# Extending the Heston Model to Forecast Motor Vehicle Collision Rates

Darren Shannon[1*], Grigorios Fountas[2]

[1] University of Limerick, Ireland
[2] Transport Research Institute, Edinburgh Napier University, UK
[*] Corresponding author: darren.shannon@ul.ie

## Abstract

We present an alternative approach to the forecasting of motor vehicle collision rates. We adopt an oft-used tool in mathematical finance, the Heston Stochastic Volatility model, to forecast the short-term and long-term evolution of motor vehicle collision rates. We incorporate a number of extensions to the Heston model to make it fit for modelling motor vehicle collision rates. We incorporate the temporally-unstable and non-deterministic nature of collision rate fluctuations, and introduce a parameter to account for periods of accelerated safety. We also adjust estimates to account for the seasonality of collision patterns. Using these parameters, we perform a short-term forecast of collision rates and explore a number of plausible scenarios using long-term forecasts. The short-term forecast shows a close affinity with realised rates (over 95% accuracy), and outperforms forecasting models currently used in road safety research (Vasicek, SARIMA, SARIMA-GARCH). The long-term scenarios suggest that modest targets to reduce collision rates (1.83% annually) and targets to reduce the fluctuations of month-to-month collision rates (by half) could have significant benefits for road safety. The median forecast in this scenario suggests a 50% fall in collision rates, with 75% of simulations suggesting that an effective change in collision rates is observed before 2044. The main benefit the model provides is eschewing the necessity for setting unreasonable safety targets that are often missed. Instead, the model presents the effects that modest and achievable targets can have on road safety over the long run, while incorporating random variability. Examining the parameters that underlie expected collision rates will aid policymakers in determining the effectiveness of implemented policies.

# 1 Introduction

The future of road safety is uncertain. Despite the push toward increasing road safety in Europe (European Commission 2018b, a, European Transport Safety Council 2020b), the fact remains that motor vehicle collisions are one of the leading causes of death both worldwide and in Europe. The prospect of connected and autonomous vehicles has revised expectations for collision and casualty rates going forward (Litman 2020). In the near-term, advanced driver assistance systems (ADAS) are expected to result in an appreciable reduction in collision rates (Yue *et al.* 2019, Shannon *et al.* 2020). Despite these advancements, motor vehicle collisions will remain a highly random and non-deterministic process. This study introduces a forecasting tool to embrace the non-determinism of this uncertainty, and provide reasonable predictions for setting and evaluating safety targets.

The tool that is introduced is an extended version of the Heston Stochastic Volatility model (Heston 1993). The Heston model is commonplace in mathematical finance. It is favoured as it leverages the evolution of two interconnected yet randomly-varying processes to investigate how an asset price may change over time. Similar assumptions can be applied to motor vehicle collisions. Evidence exists of collision rates fluctuating randomly in tandem with changing travel patterns and distances, largely due to driving exposure (Regev *et al.* 2018). In Ireland, the number of registered vehicles on road networks is trending upwards (Road Safety Authority 2019). Without concerted efforts to reduce overall collision rates, an increased number of registered vehicles and expectations of increased mobility patterns (Kröger *et al.* 2019) will lead to an inevitable increase in collision frequency. This trend will manifest while remaining tied to random fluctuations– akin to a 'random walk' with drift.

A reduction in fatality and injury risk, as well as the safe introduction of cooperative, connected and automated mobility (CCAM) solutions, is the goal of leading safety organisations worldwide. Road safety campaigns often centre on reducing aggressive driving behaviours, reducing driving while cognitively impaired, and ensuring the vehicle's safety-critical functionality is well maintained. An increasing number of campaigns also directly target younger or more inexperienced drivers who may underestimate their level of risk or become distracted while driving. These campaigns are often set in motion as a means to meet ambitious targets for collision rate reductions. Although a level of risk reduction is often achieved, the ambitious targets are often missed (World Health Organization 2018, European Transport Safety Council 2020b).

The forecasts we propose in this study can be used as an additional tool to enhance the accuracy of short-term predictions on collision rates, and analyse long-term safety target proposals. The objectives of introducing this method for long-term forecasting are two-fold. Firstly, it allows reasonable targets to be set that can be reliably met, and secondly, it allows the compounding effects of safety initiatives to be assessed through scenario analyses. A compounding effect in this regard is the assumption that a reduction in collision rates will encourage other road users to adopt safer behaviours. This will allow for even greater safety benefits to be realised in successive time periods. Using an Extended Heston model, we provide reasonable estimations for the future of road safety, without losing touch of the random variations that drive motor vehicle collision rates.

The forecasts we introduce account for the randomness of collisions by modelling collision rates using stochastic processes rather than linear equations. Developments within road safety literature have led to a sceptical view on conventional linear models that assume a fixed structure of variation in collision frequencies and severities (Xie *et al.* 2007, Anastasopoulos *et al.* 2012a, Commandeur *et al.* 2013). Collision frequencies are assumed to be non-constant and randomly-varying over time (Malyshkina *et al.* 2009, Anastasopoulos *et al.* 2012a, Mannering 2018). Collision severities, meanwhile, are assumed to be afflicted by systematic unobserved heterogeneity (Anastasopoulos *et al.* 2012b, Mannering *et al.* 2016). The assumption of unobserved heterogeneity is that similar collision events may nevertheless lead to vastly different severity outcomes. This may manifest as a result of underlying influential factors that cannot be captured in real-time statistics (Fountas *et al.* 2018, Fountas *et al.* 2019), or as a result of the non-constant effects of influential factors that can be captured in real-time statistics (Fountas *et al.* 2020, Islam *et al.* 2020). Furthermore, not only do the collision rates seem to follow a random process, but the extent of the fluctuations in collision rates have also been shown to be non-constant. Instead, it changes over time (Xie *et al.* 2007). The model presented herein sidesteps issues related to the assumption of constant variance and rates by instead assuming variance and rates evolve independently and stochastically over time. In addition to being a viable model for forecasting short-term collision rates, it can also be used to forecast long-term collision rates. It can be used for long-term forecasts as informed beliefs regarding the future variance can be used to guide the level of stochasticity within the model simulations.

We extend the Heston model to make it fit for purpose for modelling collision rates. The Extended Heston model is well-suited for interrogation in the road safety domain given the emphasis that has been placed on modelling motor vehicle collisions as random events. The platform provides a reasonable forecast of evolving collision rates, based on a combination of the latest collision statistics available and the potential safety benefits of modest rate reduction targets. Each simulation is independent. After a large number of simulations have been generated (5,000 in this study), the median rate at each time step is taken to represent the forecasted value going forward. The prediction intervals associated with these simulations are also reported.

The conventional Heston model provides three primary benefits for road safety predictions. Firstly, it does not assume a constant collision rate process. Instead, changes in collision rates are a combination of the overall trend in safety, combined with a 'random walk' whose effect is tied to a variance term. Secondly, it does not assume constant variance when calculating the extent of the 'random walk' in collision rates. Thirdly, it does not assume that the stochastic variance is centred on a fixed value throughout the simulation. Instead, two variance parameters are included – the instantaneous (current) variance, and the expected long-run variance. The long-run variance is distinct from the instantaneous variance. As such, the modeller retains control over the level of heterogeneity in the predictions. Manipulating the long-run variance offers the opportunity to incorporate expectations regarding the evolution of cooperative, connected and automated mobility (CCAM) solutions in the road environment. Namely, this variable can account for future fluctuations in collision rates due to changing traffic patterns, road designs, vehicle safety capabilities, and vehicle communication technologies. Therefore, the anticipated benefits of ADAS-enabled vehicles (Yue *et al.* 2019), highly-automated or autonomous vehicles (Cicchino 2017, Litman 2020), or the expected safety benefits associated with an overhauled road network infrastructure (Meyer *et al.* 2017, Cohen and Cavoli 2019), can be incorporated in to the model.

The adapted model provides two further benefits to the conventional model. Seasonality is commonplace in road safety literature (Malyshkina *et al.* 2009, Commandeur *et al.* 2013). We therefore incorporate a seasonal adjustment to mimic variations in observed rates. The second benefit comes from an introduction of a positive 'shock' element to incorporate accelerated periods of safety. This may be due to the introduction of safety campaigns, safety regulations, upgraded road infrastructures, safety-optimised vehicles, or integrated safety systems (such as Vision Zero (European Commission 2019a)). Despite our desire for compounded safety, we realise that the compounding effect of safety targets may not be a plausible prospect. An offset may be required in the forecasts to account for changes in human behaviour. This is in keeping with the theory of risk homeostasis, or risk compensation (Winston *et al.* 2006, Chen *et al.* 2017, Fountas *et al.* 2020, Oviedo-Trespalacios *et al.* 2020). This theory posits that an equilibrium is often met between road safety initiatives and changes in driving behaviour. In other words, as their level of absolute risk reduces, drivers may adapt their behaviour over time to become increasingly risk-seeking. Therefore, the full safety benefits of the initiative are partially offset. For example, there is evidence that early adopters of airbags and anti-lock brakes were incentivised to drive with higher levels of intensity (Winston *et al.* 2006).

Taken in full, the parameters in the Extended Heston model are latent representations of the well-established contributing factors that influence collision rates and collision severities, whether they be roadway characteristics, infrastructure elements, driver behavioural patterns, or environmental conditions. To demonstrate the application of the extended Heston model, we perform two forecasts. First, we present a 5-year forecast for 2014-2018 collision rates using model parameters discerned from 2009-2013 collision rates. Following this, we present a number of long-term forecasts detailing how collision rates may evolve from 2019-2044, based on 2014-2018 collision rates. These forecasts are based on plausible scenarios detailed from prior literature.

We note that this study is not an attempt to definitively assume what motor vehicle collision rates may look like in the future. For example, the 2009-2013 time period was characterised by Ireland's fall into the Great Recession and subsequent Eurozone debt crisis. Periods of economic instability have previously been shown to alter collision and fatality rates relative to periods of economic stability (Behnood and Mannering 2016). The 2009-2013 period of recession and debt crisis led to a downward trend in Ireland's vehicle ownership statistics (Appendix B). Although we focus on collision rates rather than collision numbers in this study, fewer vehicles on transport routes may alter collision rate dynamics. As such, predictions made on the basis of this time period may not reflect the collision rate dynamics that existed during the economic recovery period of 2014-2018, when vehicle ownership statistics trended upwards. Nevertheless, the model is introduced in this study serves as a useful platform upon which to combine randomly-evolving parameters with informed beliefs to predict scenarios relating to an uncertain future.

Section 2 describes the data used as part of the study and the patterns exhibited in Irish motor vehicle collision statistics. Section 3 describes how the Heston model is transformed from a financial model in to a road safety analysis model, and describes how the extended Heston model was formed. Section 4 presents the results of the forecasting, both short- and long-term. This section also investigates the influence exhibited by changing parameter values to investigate specific scenario analyses. Section 5 includes a discussion on how the Extended Heston model compares against other forecasting models and fits in with traditional road safety and traffic dynamic literature, while proposing further extensions that can be incorporated in follow-up studies. Section 6 concludes.

# 2 Data Description

We draw our data from the collision reports released by the Road Safety Authority (RSA) – Ireland's national road safety organisation. The reports extend from 2004-2018. Each report contains a monthly breakdown of the material damage collisions, injury collisions, and fatality collisions for the prevailing year. The progression of time accords with a consistent rise in the number of legally-registered vehicles on Irish roads.

To ensure a fair comparison of road safety over the years, the statistic we use is the collision rate after adjusting for the number of legally-registered vehicles in Ireland. To derive this rate, we divide the number of collisions for each month by the number of legally-registered vehicles in the prevailing calendar year[1]. Therefore, we report collision rates rather than absolute numbers. This provides a normalised representation of road safety over time.

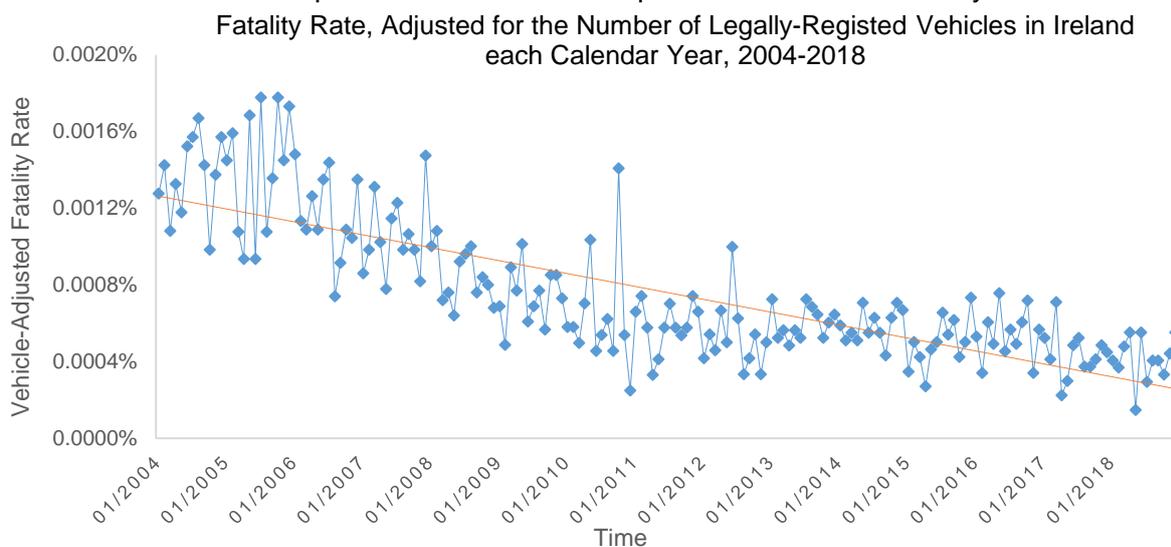

**Figure 1:** Effective road safety campaigns, improved road infrastructures, and safer vehicles has seen the fatality rate fall by almost 70% between 2004-2018

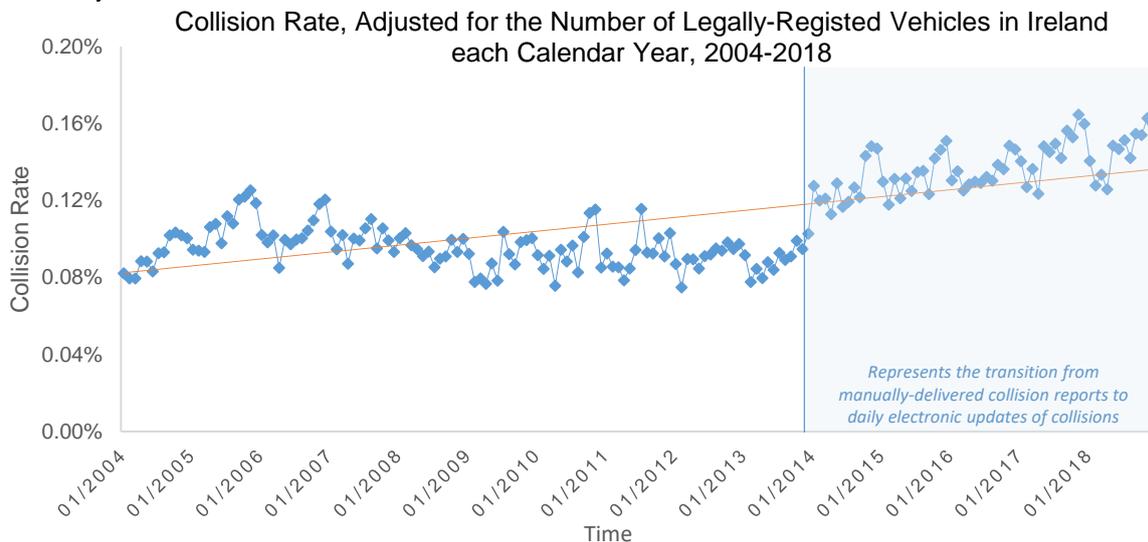

**Figure 2:** Despite the drop in fatalities in Figure 1, overall collision rates have been rising since 2014, coinciding with the switch to a digital recording of daily collision rates.

---

[1] To fully factor in traffic exposure's relationship with collision rates, the use of vehicle-miles travelled (VMT) was also investigated. However, a lack of suitable data for Ireland meant the collision rate is based on the number of legally-registered vehicles, which can also serve as a measure of exposure.

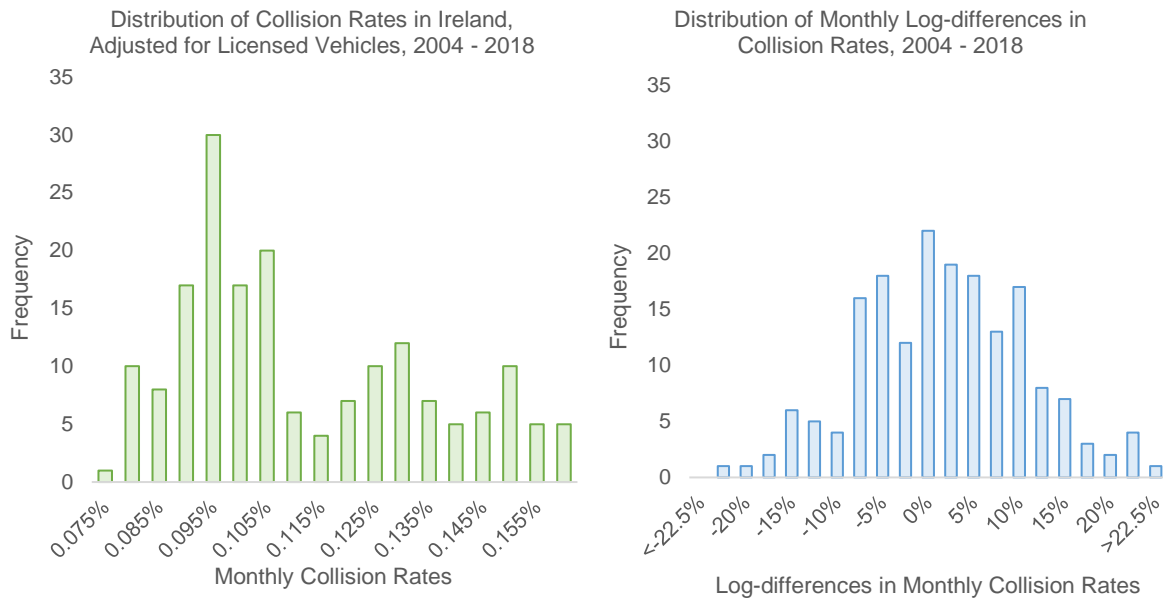

**Figure 3:** Distribution of the rate of collisions per registered vehicle on a monthly basis in Ireland between 2004-2018, and the distribution of log-differences between monthly rates.

The RSA is responsible for the collation of data collected by the national police force (An Garda Síochána). The RSA have designed and implemented effective safety campaigns that have significantly reduced the rate of fatalities on Irish roads (Figure 1). This has resulted in Ireland being ranked as the 6th safest nation in Europe for road safety (Road Safety Authority 2019). Nevertheless, total collision rates (inclusive of both material damage and casualty collisions) have been steadily rising over the last 15 years (Figure 2). Figure 3 shows that the distribution of the aggregate collision rates are log-normally distributed. However, while a slight skew exists, the monthly changes in collision rates does not exhibit a dispersion violating the limits assumed by the normal distribution.

Beginning in 2014, the method for data collation changed from manual hard-copy reports sent to the RSA to an electronic recording of the collision details, updated daily. The introduction of the new system saw a sudden increase in the number of material damage collisions. The sudden increase would suggest that the years prior to 2014 saw a large proportion of material damage collisions going unrecorded, or a consistent misplacement of collision reports. In any case, to mitigate the downward bias in rates introduced by these unrecorded collisions, we form our analysis in Section 4.2 on the basis of monthly collision rates extending from 2014-2018. Our time series therefore contains 60 monthly rates extracted from these reports (12 monthly rates over 5 years). Table 2 contains a breakdown of the collision rates throughout the 60 months and how they inform the model parameters of the extended Heston model in Section 3. This breakdown is supplemented with summary statistics describing the variability of the time series.

The lowest number of collisions occurred in April 2014 (2,875), while the highest number occurred in November 2018 (4,424). The average number of monthly collisions from 2014-2018 was 3,619. April 2014 also represented the lowest collision rate in the sample, with 0.113% of registered vehicles being involved in a collision. The highest monthly collision rate was November 2017 (0.165%) – this figure equates to 4,407 collisions amongst 2,676,000 registered vehicles.

There were large deviations among the monthly rates over the five years from 2014-2018, when measured on a log-difference scale (Table 2). The largest month-to-month decrease was a 14.71% fall in collision rates between December 2015-January 2016. The largest increase was a 18.25% rise in collision rates between April-May 2017. On average, there was a 0.38% rise in monthly collision rates, which is consistent with the rising trend in collisions (Figure 2). Log-percentage differences are preferred to arithmetic percentage changes so that symmetrical changes in rates can be reported as they relate to the starting value of a period. For example, a 10% increase on a log-difference scale, followed by a 10% decrease, will return the value to its starting position. This trait does not hold for absolute percentage changes that are measured arithmetically.

The 'volatility' of these changes, or the standard deviation of the fluctuations over a set time period, are reported on an annualised scale as $\sigma_s\sqrt{12}$. The sample standard deviation $\sigma_s$ is calculated using the values in Table 2 as $\sigma_s = \sqrt{\frac{\sum(x_i-\bar{x})^2}{n-1}}$. The 5-year volatility is 27.09%. Within-year volatilities are also calculated, which are used to calculate how volatile the scaled standard deviations are each year over the 5-year period. The 5-year 'volatility of volatility' measure is calculated as 28.71%, which indicates that the rate of change in collision rates does not remain constant from year-to-year. Both of these measures are incorporated in to our forecasting model in Section 3.

We also note a seasonal pattern in our time series data (Figure 4). Late winter to late spring months see significant downward deviations from the yearly average, with a trough typically occurring in April. This is followed by an increase in rates throughout the summer, with significant upward deviations from the yearly average occurring throughout the autumn months, typically peaking in November. It is possible that collision rates peak at this time of the year because of shorter daylight hours; poor lighting conditions has previously been linked with a higher accident risk (Jägerbrand and Sjöbergh 2016). This sinusoidal pattern is

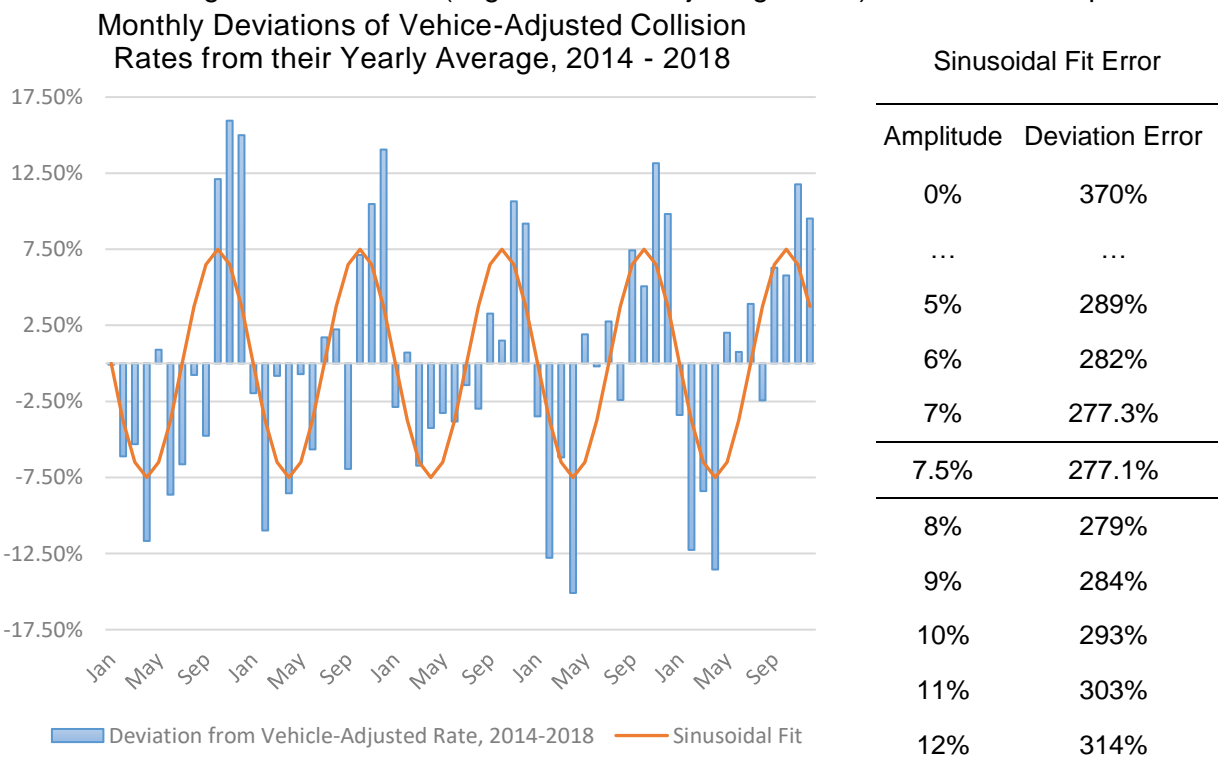

**Figure 4:** Monthly collision rates are plotted relative to the average rate of each year. The deviations of each month from their annual average follows a set pattern (lower than average the first six months, higher than average the latter six months) that can be reasoned as following that of a sine wave that begins halfway through its cycle.

Sinusoidal Fit Error

| Amplitude | Deviation Error |
|---|---|
| 0% | 370% |
| … | … |
| 5% | 289% |
| 6% | 282% |
| 7% | 277.3% |
| 7.5% | 277.1% |
| 8% | 279% |
| 9% | 284% |
| 10% | 293% |
| 11% | 303% |
| 12% | 314% |

**Table 1:** Absolute differences between sine wave cycles and monthly deviations in collision rates from their yearly average for different amplitudes.

only broken by collision patterns in August; possibly due to fewer households driving as a result of leisure breaks or a reduced school-run schedule. Nevertheless, the overall seasonality in Figure 4 generally follows that of a sine wave, which we incorporate into our model in Section 3.

The general shape of monthly collision deviations from the yearly average suggests one full sinusoidal cycle in a calendar year (angular frequency $\omega = 2\pi \times 1$). The sharp reduction in collision rates starting from January suggests that the year begins halfway through the $2\pi$ cycle, and we adjust the phase accordingly ($\varphi = \pi$). No offset is required given that the monthly deviations centre on zero each year. The amplitude $A$, or the highest peak reached during each cycle, is chosen based on a minimised error between the observable deviations and the deviations suggested by the sinusoidal cycle. The results of Table 1 suggest that an amplitude of 7.5% minimises the error, and offers a significant reduction in error over having no sinusoidal cycle incorporated ($A = 0\%$). More complex Fourier Basis functions or other Functional Data techniques may attain a closer fit to the overall seasonal pattern and the discontinuity caused by August collisions rates. However, to avoid overfitting we retain a general view on the sinusoidal fit going forward.

|  | Month | Underlying Collision Rates | | Seasonality | Volatility | | |
|---|---|---|---|---|---|---|---|
|  |  | Collisions | Collision Rate, Adjusted for Registered Vehicles (Figure 2) | Deviations from Yearly Average (Figure 3) | Monthly Collision Rate Log-differences | Yearly Volatility | Yearly Volatility Log-differences |
| Min |  | 2,875 | 0.113% | -15.09% | -14.71% |  |  |
| Max |  | 4,424 | 0.165% | 15.95% | 18.25% |  |  |
| Mean |  | 3,619 | 0.137% | 0 | 0.38% |  |  |
| Std. Dev |  | 388.95 | 0.013% | 7.69% | 7.82% |  |  |
| 2014 (Registered Vehicles: 2,546,000) | Jan | 3,252 | 0.128% | -0.09% | - | 27.38% | - |
|  | Feb | 3,056 | 0.120% | -6.11% | -6.22% |  |  |
|  | Mar | 3,082 | 0.121% | -5.31% | 0.85% |  |  |
|  | Apr | 2,875 | 0.113% | -11.67% | -6.95% |  |  |
|  | May | 3,284 | 0.129% | 0.90% | 13.30% |  |  |
|  | June | 2,974 | 0.117% | -8.63% | -9.92% |  |  |
|  | July | 3,039 | 0.119% | -6.63% | 2.16% |  |  |
|  | Aug | 3,230 | 0.127% | -0.76% | 6.10% |  |  |
|  | Sep | 3,100 | 0.122% | -4.76% | -4.11% |  |  |
|  | Oct | 3,649 | 0.143% | 12.11% | 16.31% |  |  |
|  | Nov | 3,774 | 0.148% | 15.95% | 3.37% |  |  |
|  | Dec | 3,743 | 0.147% | 15.00% | -0.82% |  |  |
| 2015 (Registered Vehicles: 2,593,000) | Jan | 3,368 | 0.130% | -1.96% | -12.39% | 31.07% | 12.63% |
|  | Feb | 3,058 | 0.118% | -10.98% | -9.66% |  |  |
|  | Mar | 3,407 | 0.131% | -0.82% | 10.81% |  |  |
|  | Apr | 3,142 | 0.121% | -8.54% | -8.10% |  |  |
|  | May | 3,411 | 0.132% | -0.71% | 8.21% |  |  |
|  | June | 3,241 | 0.125% | -5.65% | -5.11% |  |  |
|  | July | 3,494 | 0.135% | 1.71% | 7.52% |  |  |
|  | Aug | 3,512 | 0.135% | 2.23% | 0.51% |  |  |
|  | Sep | 3,197 | 0.123% | -6.94% | -9.40% |  |  |
|  | Oct | 3,680 | 0.142% | 7.12% | 14.07% |  |  |
|  | Nov | 3,795 | 0.146% | 10.47% | 3.08% |  |  |
|  | Dec | 3,918 | 0.151% | 14.05% | 3.19% |  |  |
| 2016 (Registered Vehicles: 2,645,000) | Jan | 3,450 | 0.130% | -2.86% | -14.71% | 21.47% | -36.96% |
|  | Feb | 3,577 | 0.135% | 0.71% | 3.62% |  |  |
|  | Mar | 3,313 | 0.125% | -6.72% | -7.67% |  |  |
|  | Apr | 3,401 | 0.129% | -4.24% | 2.62% |  |  |
|  | May | 3,436 | 0.130% | -3.26% | 1.02% |  |  |
|  | June | 3,416 | 0.129% | -3.82% | -0.58% |  |  |
|  | July | 3,501 | 0.132% | -1.43% | 2.46% |  |  |
|  | Aug | 3,446 | 0.130% | -2.98% | -1.58% |  |  |
|  | Sep | 3,668 | 0.139% | 3.27% | 6.24% |  |  |
|  | Oct | 3,605 | 0.136% | 1.50% | -1.73% |  |  |
|  | Nov | 3,930 | 0.149% | 10.65% | 8.63% |  |  |
|  | Dec | 3,878 | 0.147% | 9.19% | -1.33% |  |  |
| 2017 (Registered Vehicles: 2,676,000) | Jan | 3,759 | 0.140% | -3.48% | -4.28% | 29.56% | 31.98% |
|  | Feb | 3,397 | 0.127% | -12.78% | -10.13% |  |  |
|  | Mar | 3,654 | 0.137% | -6.18% | 7.29% |  |  |
|  | Apr | 3,307 | 0.124% | -15.09% | -9.98% |  |  |
|  | May | 3,969 | 0.148% | 1.91% | 18.25% |  |  |
|  | June | 3,887 | 0.145% | -0.20% | -2.09% |  |  |
|  | July | 4,002 | 0.150% | 2.76% | 2.92% |  |  |
|  | Aug | 3,801 | 0.142% | -2.40% | -5.15% |  |  |
|  | Sep | 4,184 | 0.156% | 7.43% | 9.60% |  |  |
|  | Oct | 4,092 | 0.153% | 5.07% | -2.22% |  |  |
|  | Nov | 4,407 | 0.165% | 13.15% | 7.42% |  |  |
|  | Dec | 4,277 | 0.160% | 9.82% | -2.99% |  |  |
| 2018 (Registered Vehicles: 2,718,000) | Jan | 3,824 | 0.141% | -3.39% | -12.75% | 28.41% | -3.95% |
|  | Feb | 3,473 | 0.128% | -12.26% | -9.63% |  |  |
|  | Mar | 3,626 | 0.133% | -8.39% | 4.31% |  |  |
|  | Apr | 3,422 | 0.126% | -13.55% | -5.79% |  |  |
|  | May | 4,038 | 0.149% | 2.01% | 16.55% |  |  |
|  | June | 3,988 | 0.147% | 0.75% | -1.25% |  |  |
|  | July | 4,113 | 0.151% | 3.91% | 3.09% |  |  |
|  | Aug | 3,862 | 0.142% | -2.43% | -6.30% |  |  |
|  | Sep | 4,207 | 0.155% | 6.28% | 8.56% |  |  |
|  | Oct | 4,187 | 0.154% | 5.78% | -0.48% |  |  |
|  | Nov | 4,424 | 0.163% | 11.77% | 5.51% |  |  |
|  | Dec | 4,335 | 0.159% | 9.52% | -2.03% |  |  |
| Model Parameters |  |  | December '18 Rate 0.159% | Amplitude (Table 1) 7.5% | 5-year Volatility 27.09% | 5-year Volatility of Volatility 28.71% |  |

**Table 2:** Summary statistics and model parameter derivation from a time series of monthly Irish vehicle collision data, 2014-2018.

# 3 Methodology

## 3.1 Heston model for Asset Prices

The conventional Heston Stochastic Volatility Model (Heston 1993) calculates stepwise changes in the underlying asset price $S_t$ by assuming it satisfies the stochastic differential equation

$$dS_t = \mu S_t dt + \sqrt{v_t} S_t dW_t^S \quad (1)$$

Where $v_t$, the instantaneous variance or squared volatility, is a Cox-Ingersoll-Ross (1985) process, and each change in $v_t$ is defined as

$$dv_t = \kappa(\theta - v_t)dt + \xi\sqrt{v_t}dW_t^v \quad (2)$$

Both $W_t^C$ and $W_t^v$ are Wiener processes with correlation $\rho$. Eqn. 1 and Eqn. 2 both follow a Markov Chain, where stepwise changes only depend on the current state of the process. The parameters are represented as:

- $\mu$ is the expected growth of the asset price each year. This rate is assumed to remain constant over time.
- $\theta$ is the long-run variance, or the rate to which fluctuations in the asset price will tend to over time. As $t$ tends to infinity, the instantaneous variances $v_t$ are expected to revert to $\theta$.
- $\kappa$ determines the speed at which $v_t$ reverts to $\theta$.
- $\xi$ determines the rate of change of the variance in each successive $v_t$. In addition to the assumption that asset prices follow a stochastic process, it is assumed that the extent of the fluctuations in asset prices ($\sqrt{v_t}$) are also stochastic and controlled by the constant $\xi$.

To ensure that variances $v_t$ remain positive for all $t$, the parameters are set such that $\xi^2 < 2\kappa\theta$ (the Feller Condition).

The Heston model is widely-used as it effectively models random processes that have asymmetric yet normally-distributed stepwise log-changes. It also factors in the possibility that asset price and volatility are correlated. Furthermore, it eschews the assumption that the standard deviation ($\sqrt{v_t}$) of the underlying process remains constant over time. Instead, it assumes instead that the standard deviation is a stochastic process itself, which reverts to an estimated average over time.

These traits are also applicable for modelling collision rates. Collision rates have previously been shown to be lognormally-distributed (Ma *et al.* 2015, Ma *et al.* 2016), as shown in Figure 3. Therefore, their stepwise log-differences are normally distributed (Figure 3). We additionally assume that the size of the fluctuations in collision rates are positively correlated with the collision rates. We also assume that fluctuations in collision rates are stochastic in accordance with the temporal instability assumption (Mannering 2018). This assumption states that different time periods exhibit different levels of fluctuations in collision rates. Thus, variance in collision rates is not constant over time.

## 3.2 Extended Heston model for Collision Rates

Certain extensions are made to the conventional Heston model to tailor its use for modelling monthly collision rates. The extended Heston model, for the purposes of this study, calculates stepwise changes in the underlying collision rates $C_t$ by the equation:

$$dC_t = \left(-\mu G_t C_1 dt + \sqrt{v_t} C_1 dW_t^C\right) + \left(\overline{C_t^Y} \times A \sin(2\pi ft + \varphi)\right) \quad (3)$$

Where $v_t$, the instantaneous variance or squared volatility, is a Cox-Ingersoll-Ross process, and each change in $v_t$ is defined as

$$dv_t = \kappa(\theta - v_t)dt + \xi\sqrt{v_t}dW_t^v \quad (4)$$

The $\overline{C_t^Y} \times A \sin(2\pi ft + \varphi)$ term is a adjustment that is added to account for the seasonality in collision rates (Figure 3). $\overline{C_t^Y}$ denotes the prevailing calendar-year average for simulated monthly collision rates. $A \sin(2\pi ft + \varphi)$ incorporates the sinusoidal adjustment based on the placement of $C_t$ within the calendar year. Both $W_t^C$ and $W_t^v$ remain as Wiener processes with correlation $\rho$. The level of correlation is found by computing the level of association between annual collision rates and annual collision rate volatility. The remaining parameters can be reasoned as follows:

- Rather than the expected annual growth rate, $\mu$ is set to equal the expected reduction in collision rates each year. This rate is assumed to remain constant over time, except for periods of accelerated reductions ($G_t$).
- $G_t$ represents fixed, temporary periods of accelerated reductions in collision rates. These periods represent a proxy for the safety benefits afforded by improved road infrastructures, the introduction of safety-optimised vehicles, effectively-enforced legislation, and other schemes that prove beneficial to road safety.
- $\theta$ is the long-run variance, or the rate to which month-to-month fluctuations in collision rates will tend to over time. As $t$ tends to infinity, the month-to-month variances $v_t$ are expected to revert to $\theta$.
- $\kappa$ determines the speed at which the prevailing variance of collision rates $v_t$ reverts to $\theta$.
- $\xi$ determines the extent of the fluctuations in each successive $v_t$. In addition to the assumption that collision rates follow a randomly-varying process, it is assumed that the extent of the fluctuations in month-to-month collision rates ($\sqrt{v_t}$) also vary randomly over time, and are controlled by constant $\xi$.

The extensions (Eqn. 3) to the conventional Heston model (Eqn. 1, Eqn. 2) can be summarised as:

1. including a parameter to account for periods of accelerated safety
2. incorporating an annual seasonality to the collision rates
3. adopting an generalised Wiener Process rather than an Itô Process for successive collision rates.

### 3.2.1 Accelerated Reductions

A stochastic 'rate of interest' adjustment has previously been added to the Heston model (ex: (Grzelak and Oosterlee 2011)). However, in our model we keep the 'rate of reduction' term $\mu$ constant and instead implement a short-term binary 'switch' to an accelerated rate of collision reduction. As such, we add to our predictions the plausible scenario that certain periods in the future will see an accelerated rates of collision rate reduction for short periods of time. Although similar in theory to a Markov Switching Process, we do not incorporate the Markov property; each prediction is identically and independently distributed.

Periods of accelerated collision rate reductions may follow from temporary measures (i.e. COVID-19 mobility lockdowns), or permanent measures such as updated vehicle regulations, road infrastructures optimised for safety and vehicle connectivity (such as V2X) (Johnson 2017, Najaf *et al.* 2018), or the launch of highly-automated vehicles that reduce collision rates (Litman 2020). Regulations may take any form of policy or enforcement changes that result in effective reductions in collision rates. These may include prohibiting the use of conventional vehicles on public road networks, or requiring newly-launched vehicles to have safety-optimised technology as standard (European Commission 2019b).

To incorporate these scenarios, we add a binary switching element to the constant rate of reduction parameter $\mu$. If a threshold is breached, the collision rate reduction parameter $\mu$ is expected to accelerate for $s$ periods by a multiplicative rate $G > 1$. Outside of these time periods, the multiplicative rate is set to $G = 1$. The threshold is set according to the Gompertz probability distribution function. The rate of acceleration is assumed to follow a uniform distribution bounded by $\mathbb{N}[l, h]$.

These effects last for a short period of time to account for the risk homeostasis or offset theory (Winston *et al.* 2006). The offset theory asserts that drivers tend to offset risk-reduction measures (road safety campaigns, vehicle safety technology, optimised road design, etc.) by adapting their driving behaviours to become more risk-seeking. The equilibrium reached between absolute risk-reduction and updated risk-norms means that the safety benefits of the implemented risk-reduction measures are at least somewhat offset.

To implement this accelerated-safety adjustment to the standard rate of reduction $\mu$, we draw on the Gompertz Probability Distribution Function (Figure 5). The Gompertz distribution is used in a number of different fields, for software reliability (Ohishi *et al.* 2009) and customer purchase behaviour (Bemmaor and Glady 2012), amongst others. Its use here is predicated on the assumption that effective changes in regulations, road network improvements, or vehicle technological advancements are increasingly likely to occur in the near-term (European Commission 2018b, a). However, this probability is likely to increase even further in the moderate-long term as the availability of highly-automated vehicles and safety- and communication-optimised road networks becomes increasingly feasible (Litman 2020).

The Gompertz probability distribution function is formulated as:

$$G_t(T; b, \eta) = b\eta e^{\eta} e^{\frac{bt}{T}} e^{-\eta e^{\frac{bt}{T}}} \qquad (5)$$

The chosen values for parameters $T$ (6), $b$ (0.02) and $\eta$ (0.3) results in a Gompertz probability distribution function that resembles that of a truncated normal distribution (Figure 5). Each collision rate reduction period is randomly-predicted, based on the dynamic relationship between the uniform distribution and the Gompertz distribution (Figure 5). Therefore, each simulation will project a differing number of accelerated rate reduction periods. Nevertheless, setting $T = 6$ suggests that six accelerated periods of collision rate reduction are expected (but not guaranteed) over the next century (the span of the Gompertz function). The decline in the latter half of the distribution follows the assumption that an array of changes to reduce collision rates would also reduce the likelihood of further changes in the long-term, as policy efforts and capital may be diverted elsewhere.

Based on these assumptions, we add the following adjustment to Eqn. (1):

$$dC_t = -\mu G_t C_1 dt + \sqrt{v_t} C_1 dW_t^C \qquad (6)$$

Where

$$G_{t\ldots t+s} = \begin{cases} \alpha, if\ G(T;b,\eta) > \text{unif}[0,1], \\ \quad \alpha \in \text{unif}(\mathbb{N}[l,h]) \\ 1, if\ G(T;b,\eta) < \text{unif}[0,1] \end{cases} \quad (7)$$

In this case, the acceleration multiplier $G_t$ is a Boolean operator that determines whether the stepwise rate of reduction in collision rates is accelerated at a rate of α, or remains at the standard rate. The acceleration parameter α is a uniformly-distributed integer drawn from $\text{unif}[l, h]$. Assuming that effective changes may instigate a $2x$-$5x$ accelerated reduction in collision rates lasting for a period of 36 months, this would equate to $s$ = 36 months and α would take the form $\text{unif}[2,5]$. After incorporating the acceleration parameter $\alpha = \text{unif}[2,5]$ and the Gompertz distribution $G(6; 0.03, 0.2)$ assumptions, the acceleration multiplier $G_t$ approximately equates to a mean average of $\bar{G} \cong 1.40$ across 5,000 simulations. As such, the rate of reduction rate $\mu G_t$ is approximately 40% higher than the baseline $\mu$.

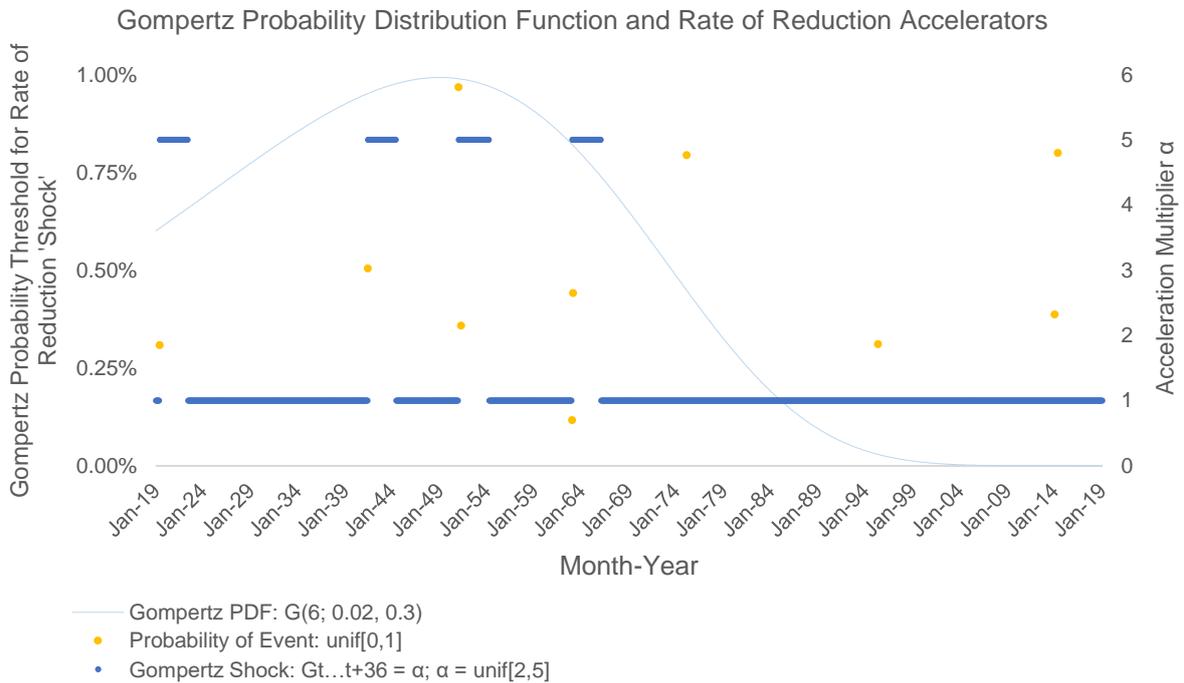

**Figure 5:** Gompertz probability distribution function overlaid with the 'acceleration in safety' multiplier α and uniformly-distributed random 'predictions' for the month in which the acceleration multiplier will rise. For each random 'prediction' (yellow dot varying between 0%-100%) that falls below the dynamic threshold set by the Gompertz curve, there will be a Gompertz 'shock', or an accelerated rate of safety for the following 36 months.

### 3.2.2 Seasonality

As noted in Section 2, the rate of motor vehicle collisions in Ireland follows a cyclical pattern (Figure 4). Spring typically sees the lowest rate of collisions, with a trough in April, while autumn months see the highest rate of collisions, typically peaking in November. To account for this seasonality, we apply a sinusoidal adjustment to the collision rates. Seasonal adjustments have previously been added to the Heston Stochastic Volatility model (Arismendi *et al.* 2016), however in our case we adjust the expected collision rate rather than the rate of fluctuation.

$$C_t = C_t + \overline{C_t^Y} \times A \sin(2\pi f t + \varphi) \quad (8)$$

Where

- $\overline{C_t^Y}$ is the bounded calendar-year average (January-December) for the prevailing collision rate.

- $A$ is the amplitude, or the highest peak reached in the seasonal cycle.
- $f$ is the frequency, or the number of sinusoidal cycles per unit of time (year).
- $2\pi f$ is the angular frequency; the rate of change of sinusoidal cycles per unit of time (year).
- $\varphi$ is the phase, or the position of the sinusoid cycle at $t = 0$.

In our long-term forecasts (Section 4.2), the seasonal adjustments to the collision rate predictions take the form:

$$C_t = C_t + \overline{C_t^Y} \times (0.075 \sin(2\pi t + \pi)) \quad (9)$$

According to Table 1 in Section 2, sinusoidal error is minimised when $A$ is 7.5%. $f$ is set to one cycle per year, while successive time steps are set to $\frac{1}{12}$ to account for monthly iterations. $\varphi$ is set to $\pi$, in order to enable the sinusoid to begin half-way through its cycle.

### 3.2.3 Independent Collision Steps, Dependent Variance Steps

The largest deviation from the conventional Heston model is a departure from the assumption of state-dependent changes in successive collision rates $C_t$. This is in contrast to the common Heston model convention that assumes asset price follows an Itô Process (Bergomi 2015). Instead, we assume that monthly collision rates follow a random walk. The state-dependent assumption asserts that successive values in the process are scaled to the size of the immediately-prior value $C_t$. In this way, if collision rates are to reach near-zero values, the state-dependent process would ensure there is a high-likelihood that only a near-zero change can occur for the following time step. In a decaying process such as that underlying an iterative 'rate of reduction' process, this has the effect of 'trapping' collision rates near-zero if a random simulation reaches this level. This is counter-intuitive in the context of monthly collision rates. Collision rates are primarily dependent on travel patterns and traffic volumes (Lassarre 2001, Bijleveld *et al.* 2010, Commandeur *et al.* 2013, Regev *et al.* 2018), but are otherwise unrelated enough to be assumed as temporally independent. As such, we scale successive values in the series only to the initial value $C_1$. Rates are further floored at zero to prevent the simulation of negative collision rates. As such, this truncation scheme has the effect of 'bouncing' predicted collision rates off zero once low levels are reached, rather than 'trapping' the series.

However, we do assume that successive changes in variances $v_t$ are time-dependent. We retain this assumption on the basis that the variance in monthly collision rates will trend toward a long-term average $\theta$ over time. As such, we retain the conjecture that our uncertainty as to the size of future variance movements is proportional to the level of the prevailing variance. We anticipate changes in variance as road infrastructures become more optimised and V2X communication becomes an increasingly feasible prospect. This may result in higher variance (reduced collision rates punctuated by higher spikes) or lower variance (optimised road networks evenly dispersing collision frequencies).

This is distinct from the assumption that an interconnected society will definitively reduce month-on-month collision rates $C_t$. Rather, it is assumed that an interconnected society will result in a reduction in the random fluctuations $(\sqrt{v_t})$ of monthly collision rates as transport networks become increasingly optimised.

# 4 Results
## 4.1 The Case for 2014-2018

As noted in Figure 3, the transition to daily updates of electronically-recorded collisions caused a spike in collision rates. Rather than significantly affecting the underlying dynamics of the collision rate process, it is believed that the transition from manual reporting to electronic recordings in January 2014 primarily affected the baseline level of collision rates. For this reason, we use the parameters derived from collision rates in 2009-2013 (Appendix B, Appendix C) to forecast estimates of the collision rates in 2014-2018. The results of the procedure (Figure 6) highlights the efficacy of the model proposed in Section 3. All statistical procedures and forecasts were completed in Matlab.

The scenario presented in Figure 6 is based on the median values of 5,000 random simulations, and is predicated on no expected changes in overall collision rates going forward ($\mu = 0$). Furthermore, month-to-month fluctuations between 2009-2013 equate to an estimated volatility of 40.6% each year (equivalent to a variance of 16.5%). We assume that the rate of change in monthly fluctuations will remain the same over the five years from 2014-2018, and there will be no accelerated periods of safety. In other words, we foresee no overhauls of transport infrastructure, driving behaviour, or vehicle technologies that will minimise the difference in month-to-month rate fluctuations from 2014-2018[2]. Hence, we assume that the current rate of fluctuations are equal to the long-run average of fluctuations ($v_0 = \theta$), and that the safety acceleration parameter will remain constant ($\bar{G} = 1$).

Although the within-year volatilities are widely-varying, the year-to-year differences are tamer, with the volatility of volatilities ($\xi$) calculated to be 23%. Given that we expect the current variance to equal the long-run variance, the mean reversion rate ($\kappa$) is set to the minimum that would satisfy the Feller Condition (16%). We also note that the correlation between annual collision rates and volatilities are 60%, indicating that higher volatilities are associated with higher collision rates and vice versa. The sinusoidal pattern that minimised monthly collision rates from 2009-2013 has a peak of 9% (Appendix C).

|         | Mean Absolute Error (MAE) | Root Mean Squared Error (RMSE) | Mean Absolute Percentage Error (MAPE) |
|---------|---------------------------|--------------------------------|---------------------------------------|
| 2014    | 0.007%                    | 0.008%                         | 5.29%                                 |
| 2015    | 0.006%                    | 0.008%                         | 4.68%                                 |
| 2016    | 0.006%                    | 0.007%                         | 4.20%                                 |
| 2017    | 0.007%                    | 0.009%                         | 4.98%                                 |
| 2018    | 0.006%                    | 0.008%                         | 4.39%                                 |
| Average | 0.006%                    | 0.008%                         | 4.71%                                 |

**Table 3:** Error statistics for Heston forecasts of monthly collision rates between 2014-2018 (5000 simulations). The absolute differences between the forecasted values and observed values are minimal. The relative (%) differences are consistent year-on-year, averaging over 95% accuracy over the 5-year time period.

The baseline rate indicates that 0.128% of registered vehicles were involved in collisions in January 2014. Starting from this rate[3], and using the above parameters, the estimates from the extended Heston model outlined in Section 3 closely aligns with observed rates for the following five years (Figure 6). It is predicted that collision rates continue to follow a sinusoidal pattern, and begin drifting upwards of the baseline from year two (2016). Monthly collision rates were forecasted to remain above the baseline from 2017-2018 and rise further – a trait that is also observable in the realised rates.

---

[2] Granted, this is easy to assume in hindsight.
[3] The move in data collection from manual reports to electronic recordings of vehicles in January 2014 dramatically increased the collision rate (Figure 2). It is believed that the January 2014 collision rate represented a 'new norm' that was not reflected in the December 2013 rate. Therefore, we started from the January 2014 level, rather than December 2013.

Error statistics measuring the absolute and relative (%) difference between the forecasted rates and the observed rates are available in Table 3. The results indicate a consistent year-on-year forecasting accuracy. Absolute inaccuracies remain less than 0.01%, while relative inaccuracies remain consistent at approximately 5% each year. As such, the forecasting accuracy remains at an approximate 95% level each year, with a 5-year monthly average forecasting accuracy of 95.3%. Prompted with these results, we use the collision parameters observable in 2014-2018 monthly collision rates, as detailed in Section 2, to forecast collision rates going forward.

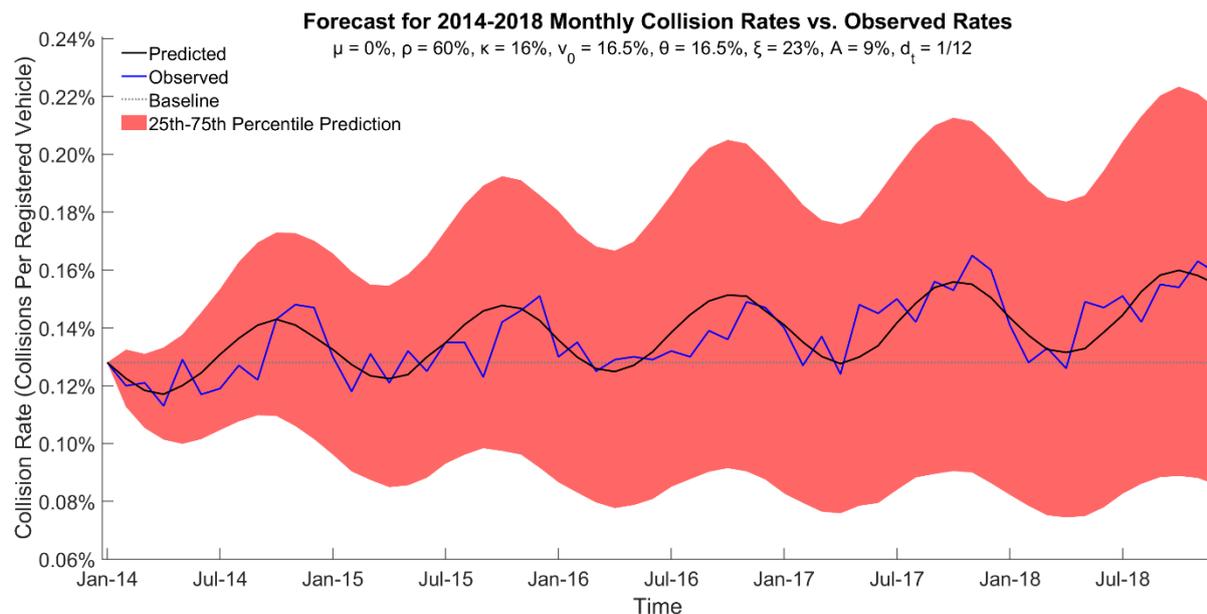

**Figure 6:** Observed vs. Predicted Values for Monthly Collision Rates from 2014-2018, starting from the January 2014 rate of 0.128%. Predictions were based on the Extended Heston model outlined in Section 3, where the parameters were discerned from monthly collision rates covering 2009-2013 (Table B1, Table C1).

### 4.2 Looking Forward: The Case for 2019-2044

We perform the same process as in Section 4.1 to forecast how the future of collision rates may evolve. We must note that the effects of COVID-19 have no doubt disrupted the overall accuracy of the results, with the EU seeing a 36% year-on-year fall in fatalities between April 2019 and April 2020 (European Transport Safety Council 2020a). Although a consequential shift in societal transport patterns as a result of COVID-19 is plausible, it remains to be seen whether the downward shift in collision rates is sustainable. Presently, we expect the trend in collision rates to return to approximate pre-COVID-19 levels. Instead, we expect for road safety initiatives to be the dominant factor in the evolution of collision rates.

As stated previously, the results generated in this study are not attempts to definitively assume what motor vehicle collision rates may look like in the future. Rather, it is an exploration of what is plausible. For this reason, our purpose in this subsection is to interrogate the parameters as they pertain to scenario analyses for road safety.

### 4.2.1 Baseline Scenario – no effective change

The baseline forecast is formed on little expected change in the underlying dynamics of overall collision rates, beyond the schemes currently implemented ($\mu = 0$). Current variance measures will also remain unchanged over time ($\theta = 7.3\%$), and no accelerated periods of safety are expected ($\bar{G} = 1$). Accelerated periods of safety in this context are assumed to be a proxy for sudden, significant developments in road safety lasting for a short period of time. Examples of these developments include a widespread assimilation of connected and autonomous vehicles, an increased market penetration of vehicles equipped with a suite of ADAS technologies, effective legislation improving the aggregate level of safety on transport routes, or upgraded transport infrastructure. For the purposes of this study, these short, fixed-term periods are assumed to last 36 months. Figure 7 highlights a long-term forecast of the model outlined in Section 3 using data described in Section 2. The parameters are formed on the collision rates from January 2014-December 2018. Also provided is a close-up view of the forecast over the coming five years, beginning from 2019[4].

In this scenario, January 2024 rates (a relatively-speaking 'safer' month) are expected to reach the level of December 2018 rates (a relatively-speaking 'dangerous' month). This signals a sharper rise from historical rates. January collision rates did not reach trailing December collision rates in any 5-year period from 2004-2014 (i.e. prior to the switch to electronic recordings).

The median expectations are that collision rates will remain relatively constant until 2028, at which point there will be shift to a slight upward trend until 2037. After 2036, it is expected that collision rates will trend upwards at an accelerating rate. By 2044, without no further market interventions to reduce the collision rate, the median expectation is that collision rates will rise by approximately 25% to 0.20%. This is bounded by a 50% prediction interval of 0.093% to 0.32%, and an 80% prediction interval of 0.034% to 0.45%.

However, the likelihood of road safety remaining as it is at present is low. Technological advancements, upgraded road infrastructures, and an accelerated focus on reducing road fatalities are expected to change the dynamics of driving. In the near-term, ADAS-enabled vehicles are expected to have an appreciable impact in reducing collision rates (Yue *et al.* 2019). In the long-term, highly-automated and fully-autonomous vehicles are expected to significantly reduce the frequency of collisions (Cicchino 2017, Litman 2020). However, it is expected that this will be somewhat offset by an expected increase in the number of miles travelled by each vehicle (Fagnant and Kockelman 2015, Clements and Kockelman 2017). Regardless of the effects that these advancements will have for road safety, it can be assumed that the collision rate may trend downwards and there will be a deviation from current collision rate fluctuations. We explore these scenarios in Section 4.2.2.

---

[4] At the time of writing, the last publicly-released monthly collision rate in Ireland was for December 2018.

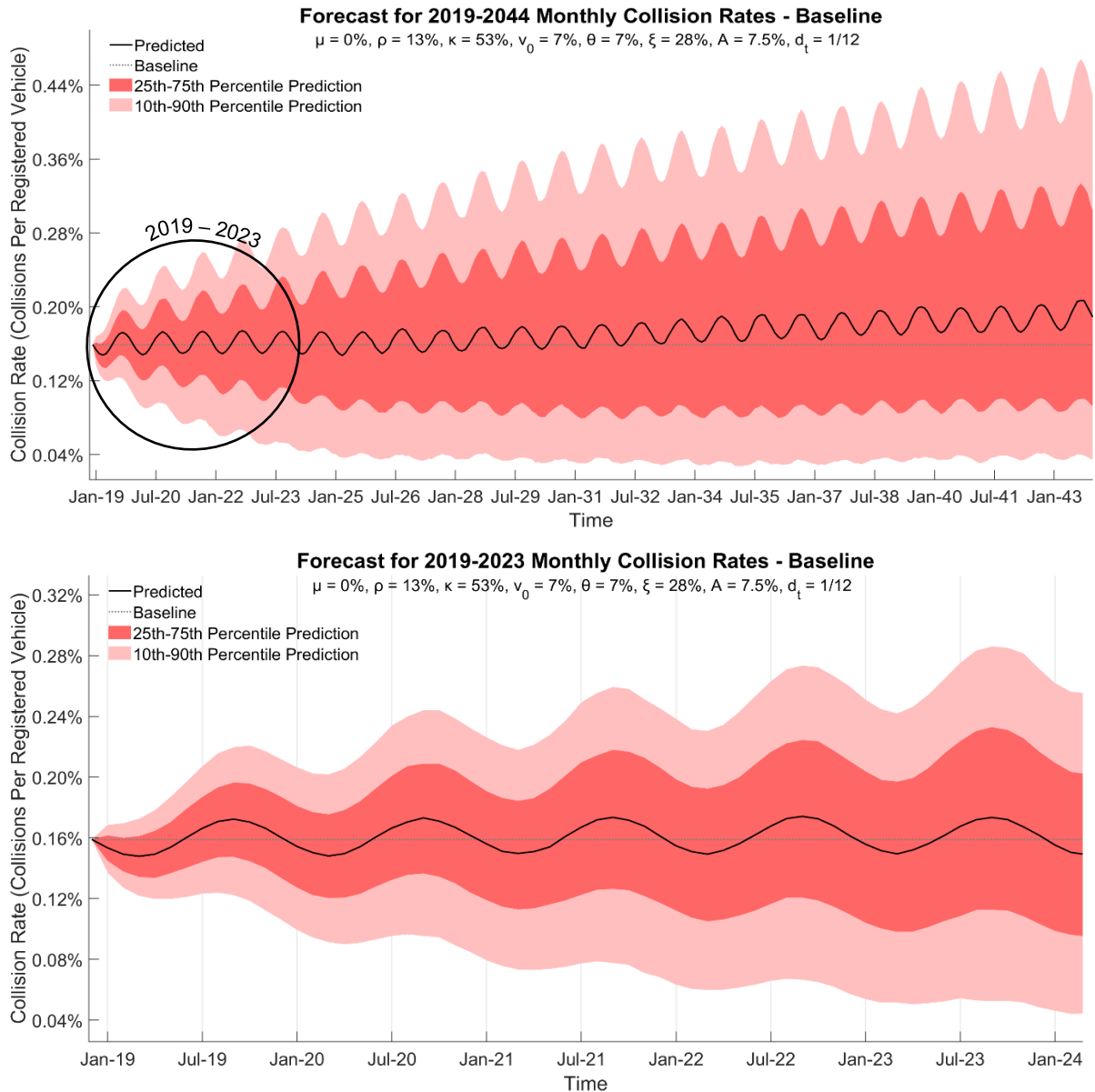

**Figure 7**: Forecasted Monthly Collision Rates from 2019-2044, starting from the December 2018 rate of 0.159%. Also provided is a close-up of the forecast, covering 2019-2023. Forecasts are based on the model outlined in Section 3, where the parameters were derived from monthly collision rates covering 2014-2018 (Table 1, Table 2)

### 4.2.2 Parameter Effects – Scenario Analyses

A number of scenarios for the future of road safety are plausible. Some are based on a reduction in collision rates (Arbib and Seba 2017, Cicchino 2017). Some others suggest a conflicting relationship between enhanced safety and risk exposure as a result of a change in miles travelled per vehicle (Fagnant and Kockelman 2015, Clements and Kockelman 2017, Litman 2020). A more cautious view posits that higher levels of vehicle automation may increase the relative risk associated with novice drivers, given that more active driver monitoring systems may encourage distracted driving or driving behaviours synonymous with inexperience (Jannusch *et al.* 2021). A further suggestion is that effective planning is key to enhancing urban transport safety for the next generation of vehicles (Meyer *et al.* 2017, Cohen and Cavoli 2019, Litman 2020). We detail each of these in turn, as they relate to the extended Heston model. The baseline parameters have been discerned in Section 2. Based on monthly collision rates from 2014-2018, the volatility of these rates was determined to be 27.09%, corresponding to a variance of $v_0 = 7.3\%$.

We structure the parameters such that the model mimics the scenarios presented in the aforementioned studies, and analyse the results. Some assume that no deviations in collision rates was expected ($\mu$ = 0%). Some further assume that the current variance will remain the same over the long-run ($v_0 = \theta$ = 7.3%), and no accelerated periods of safety will occur going forward ($\bar{G}$ = 1). The expected collision rates associated with each scenario are provided in Figure 8, and summarised in Table 4. Also provided are prediction percentiles.

In scenarios 1 and 2, we incorporate into our forecasts an assumption that there will be no short periods of accelerated rates of reduction ($\bar{G}$ = 1). In other words, these scenarios explore the evolution of collision rates in an environment where steps to optimise road safety and road infrastructure do not have a sizeable impact. These impacts may manifest as encouraging changes in driving behaviour, vehicle ownership, or traffic patterns, which may lead to gradual rather than accelerated changes in collision rates. These scenarios are also based on maintaining current levels of variance ($v_0 = \theta$). Scenario 1 assumes that no 'collision rate reduction' target will be set ($\mu$ = 0). In scenario 2, we assume an annual collision rate reduction target that is set equal to the number of newly-registered vehicles each year ($\mu$ = 1.83%)[5].

Scenario 1 is equivalent to the baseline forecast in Section 4.2.1 that forecasts a steady increase in collision rates. Scenario 2 highlights the importance of the accelerated periods of safety. This scenario suggests that rate reduction targets would not be sustainable unless partnered with changes in driving behaviour, vehicle ownership, traffic patterns, or effective legislation. The results suggest that collision rate reduction targets will initially be realised. However, a consistent variance in fluctuations means that these reductions eventually stabilise, and collision rates begin to revert to the initial rate toward the end of the forecast.

Scenarios 3 and 4 are based on the assumption that the variance in monthly collision rates will increase over time ($v_0 < \theta$). In other words, the current collision rate cycle will consolidate. Fewer collisions will be expected from late winter to late spring, but more collisions will be expected from mid-autumn to mid-winter. The effect will be such that rates will exhibit a larger deviation from the yearly average, increasing the fluctuations in monthly collision rates. As with scenarios 1 and 2, scenarios 3 and 4 assume no downward trend in collision rates ($\mu$ = 0) and a modest downward trend in collision rates ($\mu$ = 1.83%), respectively.

Scenario 3 may present in an environment where no infrastructural changes have been made and traffic patterns consolidate on current peak-travel times. Assuming an increasing number of vehicles on transport routes, higher peaks and lower troughs of traffic volume would increase the level of variance in collision rates. The extended Heston model suggests that in this scenario, the monthly collision rate will rise 50% from its initial rate by 2044, to 0.24%. However, regulatory planning suggests that this scenario is the most unlikely of the six presented. Progress has been already made on introducing further vehicle legislations (European Commission 2018b, a, 2019b), and transforming urban areas into sustainable transport routes. The European Commission have proposed a mission to transform European cities into smart cities that encourage safe mobility for all modes of transport (European Commission 2020), while the US Department of Transportation have taken steps to fund 'smart city' developments (US DOT 2016).

---

[5] Chosen to be a modest yet relatable target. The average year-on-year increase in licensed vehicles for the period 2014-2018 was 1.83% (Table 2). Hence, our initial road safety targets would be to offset this figure.

## No Change in Collision Rates ($\mu = 0\%, \nu_0 = 7.3\%$)

## Fall in Collision Rates ($\mu = 1.8\%, \nu_0 = 7.3\%$)

### Constant Variance, No Periods of Accelerated Safety ($\theta = 7.3\%, \bar{G} = 1$)

**1.**
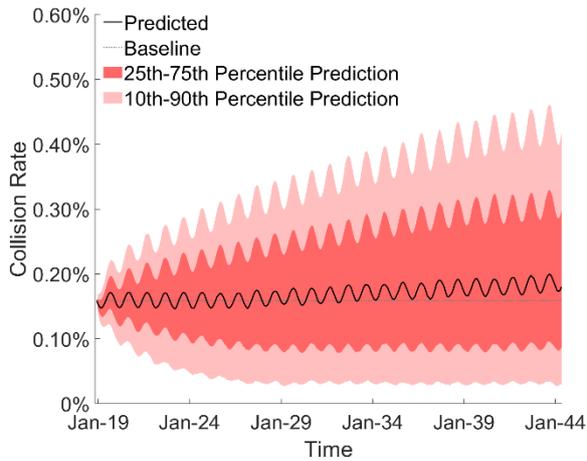

**2.**
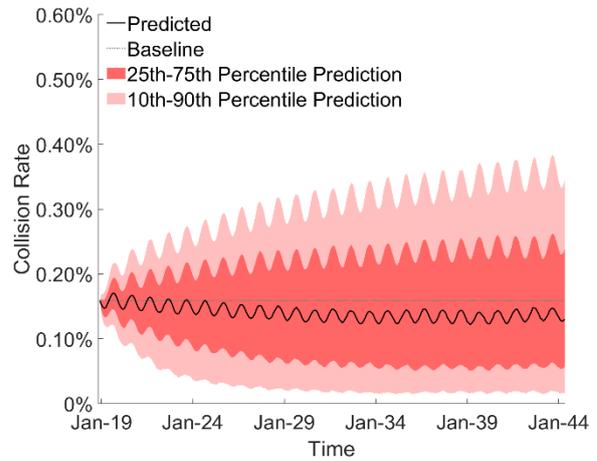

### Rise in Long-run Variance, Periods of Accelerated Safety Expected ($\theta = 14.6\%, \bar{G} > 1$)

**3.**
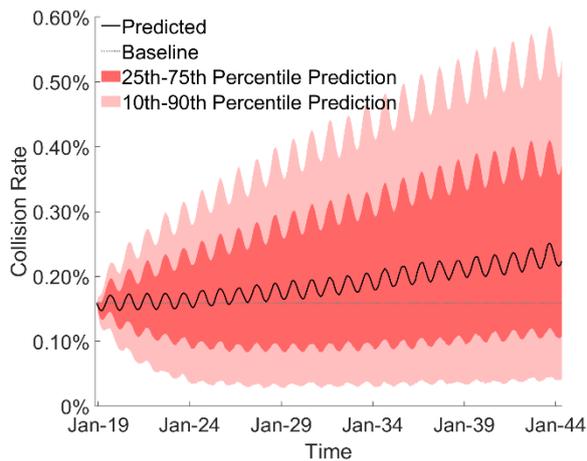

**4.**
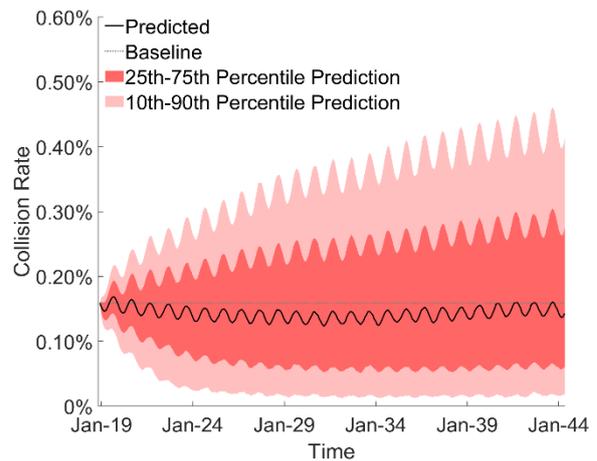

### Fall in Long-run Variance, Periods of Accelerated Safety Expected ($\theta = 3.7\%, \bar{G} > 1$)

**5.**
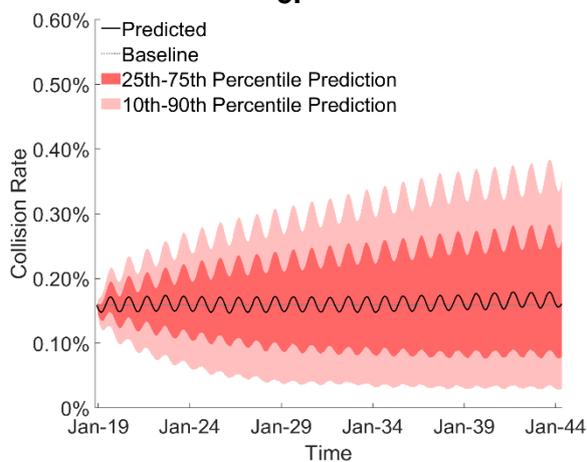

**6.**
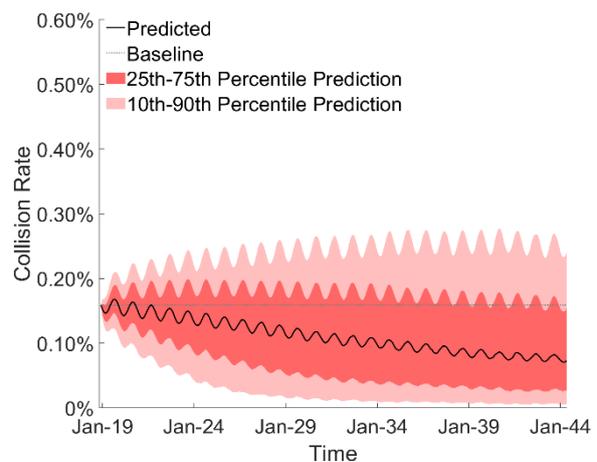

**Figure 8:** Model results for six plausible scenarios forecasting the evolution of motor vehicle collision rates from 2019-2044, using the extended Heston model. Annual 'rate reduction' target rates combined with gradual falls in variance (scenario 6) is the only scenario that suggests a persistent drop in collision rates. Note also that scenario 1 describes the 'Baseline' scenario from Section 4.2.1.

Scenario 4 indicates an environment where modest drops in collision rates are likely to arrive in conjunction with larger fluctuations. The model suggests that, despite implemented efforts to reduce collision rates, increasing the fluctuations would negate the effectiveness of these campaigns. This scenario is most likely in an environment where road safety is a priority, but has been offset by counteractive patterns in driving behaviours. An increased variance in collision rates is therefore plausible. Miles travelled per vehicle is expected to increase over the next couple of generations (Fagnant and Kockelman 2015, Clements and Kockelman 2017, Litman 2020). These counteractive patterns may include a rise in risk-seeking behaviours, significant increases in vehicle-miles travelled, or greater-than-expected levels of malfunctions in vehicle software or hardware. Young novice drivers may also pose a greater relative risk, as the increased responsibility being placed on driver monitoring systems may encourage distracted driving (Jannusch *et al.* 2021). While absolute collision frequencies may fall, the risk exposure of vehicles may increase, leading to a higher probability of periods with greater fluctuations in collision rates than those that are seen today. As with scenario 2, developments in scenario 4 would lead to initial reductions in collision rates. Unlike scenario 2, however, collision rates in this scenario are forecast to revert to the baseline rate (0.16%) by 2044.

| | Summary of Extended Heston Scenario Simulations (2019-2044) | | | |
|---|---|---|---|---|
| | Expected Collision Rate Reductions ($\mu$) | Expected Collision Rate Variability ($\Delta\theta$) | Expectation of Accelerated Periods of Safety ($G$) | Outcome |
| Scenario 1 | No drop in trend | No change in variability | No accelerated periods of safety | Slight increases in collision rates |
| Scenario 2 | Expected drop in trend | No change in variability | No accelerated periods of safety | Slight reductions in collision rates |
| Scenario 3 | No drop in trend | Increase in variability | Periods of accelerated safety | Significant increases in collision rates |
| Scenario 4 | Expected drop in trend | Increase in variability | Periods of accelerated safety | Slight reductions in collision rates (medium-term), no appreciable difference in collision rates (long-term) |
| Scenario 5 | No drop in trend | Drop in variability | Periods of accelerated safety | No appreciable difference in collision rates |
| Scenario 6 | Expected drop in trend | Drop in variability | Periods of accelerated safety | Significant decline in collision rates |

**Table 4:** A summary of the scenarios outlined in Figure 8, and the associated outcomes in terms of expected changes to collision rates over the period 2019-2044, relative to current levels.

Scenarios 5 and 6 are based on the assumption that the variance in monthly collision rates will decrease over time ($v_0 > \theta$). A lower variance is indicative of environments where month-to-month collision rate fluctuations taper over time. This may result from behavioural changes in driving or road safety developments. As before, scenarios 5 and 6 assume no downward trend in collision rates ($\mu$ = 0) and a modest downward trend in collision rates ($\mu$ = 1.83%), respectively. Scenario 5 may occur in an environment where the general make-up of traffic, vehicle type and road infrastructure does not change. As such, this scenario is predicated on few widespread 'rate reduction' initiatives being introduced. Rather, the largest change in this scenario comes from alterations in driving or travel patterns. Given that motor vehicle collisions are largely tied to the volume of traffic (Lassarre 2001, Bijleveld *et al.* 2010, Commandeur *et al.* 2013, Regev *et al.* 2018), this scenario may present as a move away from the 'peak' and 'trough' traffic cycles of autumn-winter and spring-summer. More dispersed traffic may result in the same relative number of incidents, but the seasonal fluctuations will be tamer. In this scenario, the results of the extended Heston model suggests that a reduction in variance, without a corresponding reduction in expected collision rates, has a stagnating effect on collision rates. As such, rates remain relatively constant for a significant period of time.

Scenario 6 describes a scenario where collision rate targets coincide with smaller fluctuations in collision rates. This scenario may present where significant efforts have been made to cater for safer transport routes and road networks contain a sizeable share of ADAS-enabled vehicles, as well as connected and autonomous vehicles. These adaptions may take the form of high market penetration rates, backed with updated regulations and upgraded road infrastructures that are optimised for road users' safety. The model suggests that an environment containing road networks capable of supporting safer transport routes and safety-optimised vehicles can significantly reduce the number of 'preventable' incidents (Cicchino 2017). As such, month-to-month fluctuations would be expected to taper over time. The median prediction suggests a 50% fall in collision rates (from 0.159% to 0.08%), before beginning to plateau. These predictions are reasonable. Ireland witnessed a 57% fall in road fatalities between January 2004 and December 2018 (Figure 1), despite the rise in the number of legally-registered vehicles (Table 2), while neighbouring Scotland witnessed a 61% fall in collisions over 23 years despite a 27% rise in traffic (Transport Scotland 2019). 75% of simulations suggest that some sort of effective change is predicted by 2044.

The suggestion that collision rates plateau over time is not implausible. This scenario is likely in an environment where autonomous or highly-automated vehicles are commonplace in 15-25 years, and a prominent use for vehicles is 'mobility-as-a-service' (MaaS) (US DOT 2017, Kaltenhäuser *et al.* 2020). It is anticipated that falling costs associated with ride-sharing services or MaaS will decrease the utility of vehicle ownership, leading to fewer privately-owned vehicles (Ho *et al.* 2018, Kaltenhäuser *et al.* 2020, Litman 2020). An expected fall in collisions, in conjunction with a fall in vehicle ownership rates, may stabilise collision rates when a high market share is achieved.

# 5 Discussion
## 5.1 Comparison of Extended Heston vs. Alternative Forecasting Methods

Forecasting methods have previously been employed in road safety and traffic flow dynamics literature – most notably variants of ARIMA models (Ramstedt 2008), GARCH models (Zhang *et al.* 2013), or combinations of both (Chen *et al.* 2011, Guo *et al.* 2014). SARIMA models have also been used to forecast road safety data that exhibited seasonality (Zhang *et al.* 2015). Accurate forecasts on the stochastic nature of traffic flows were also achieved through the use of the Vasicek model, a stochastic process that has long been used for interest-rate modelling (Rajabzadeh *et al.* 2017).

To demonstrate the efficacy of the Extended Heston model in forecasting short-term collision rates, the same process described in Section 4 is applied to the Vasicek model used in Rajabzadeh *et al.* (2017), and the SARIMA model used in Zhang *et al.* (2015). SARIMA-GARCH models were also generated, as per Chen *et al.* (2011), Guo *et al.* (2014), but demonstrated no appreciable difference to the forecasts produced by the SARIMA model alone. Hence, they are not considered fully in Table 5. Full details on the performance of the adjusted Vasicek model and the SARIMA model are available in Appendix A. However, Table 5 demonstrates that the Extended Heston model provides a superior performance than the other models in out-of-sample forecasting accuracy tests. The forecast associated with the Extended Heston model produced the closest affinity to 2014-2018 collision rates, based on 2009-2013 parameters.

|         | Mean Absolute Percentage Error (MAPE) | | |
|---------|---------------------------------------|---|---|
|         | Extended Heston *(from Table 3)* | Adjusted Vasicek[6] | SARIMA[7] *Optimal fit:* $(7,1,1) \times (1,1,2)_{12}$ |
| 2014    | 5.29% | 5.55% | 4.47% |
| 2015    | 4.68% | 5.62% | 4.43% |
| 2016    | 4.20% | 4.95% | 4.99% |
| 2017    | 4.98% | 9.28% | 11.91% |
| 2018    | 4.39% | 9.86% | 11.11% |
| Average | 4.71% | 7.05% | 7.38% |

**Table 5:** Error statistics for Extended Heston forecasts (5000 simulations), Adjusted Vasicek forecasts (5000 simulations), and a SARIMA$(7,1,1) \times (1,1,2)_{12}$ model forecast, respectively. The error statistics describe the relative (%) differences between forecasted and observed monthly collision rates between 2014-2018 for each model. The Extended Heston model proved the most consistent year-on-year, as well as averaging over 95% accuracy over the 5-year time period.

## 5.2 Stochastic Volatility as a Forecasting Tool

Stochastic Volatility models are a parsimonious and non-deterministic means of forecasting uncertainty, and are underutilised in road safety research. However, they have been employed to express the temporally-changing dynamics of traffic volume patterns (Tsekeris and Stathopoulos 2010, Rajabzadeh *et al.* 2017). The latter study highlights the benefits that can be gleaned from cross-disciplinary perspectives between finance and traffic dynamics.

In this study, we introduce the Heston model as an alternative means of forecasting scenarios for road safety rather than traffic volumes. Although we demonstrate its efficacy in producing short-term forecasts in Section 4.1 and Section 5.1, the predominant value associated with the Extended Heston model lies in the flexibility in its parameters. The benefit this provides is that a combination of observed data and reasonable expectations can be incorporated into the parameters. Adjusting these parameters based on expert insight provides an avenue for analytically discussing the expected evolution of road safety through scenario analyses. Firstly, the extent of the stochastic movements in the variance can be controlled using the 'volatility of volatility' ($\xi$) term. Secondly, it incorporates the assumption that excessive swings in collision rates is associated with the absolute size of collision rates. The extent of this association can be controlled using the correlation parameter ($\rho$). Furthermore, the model can account for the likely prospect that the dynamics of the underlying process may change over time. The instantaneous variance ($v_t$), despite its stochastic nature, can be set to trend toward a long-run variance ($\theta$), the rate of change of which is controlled by the long-run reversion term ($\kappa$). As such, it is possible to factor in beliefs regarding the long-term variance of collision rates, which have been shown to vary in different time periods. These variables can be combined with volatility estimates gleaned from realised collision rates to provide accurate short-term forecasts and reasonable long-term forecasts for collision rates, as shown in Section 4.1, Section 4.2.2, and Section 5.1 respectively.

---

[6] To ensure comparability, the adjustments proposed for the Extended Heston model (Section 3.2) are applied to the Vasicek model to create an 'adjusted Vasicek model'. These adjustments are detailed in Appendix A.
[7] In addition to the SARIMA model, a SARIMA-GARCH model was considered. The GARCH model introduces heteroscedastic innovations to the SARIMA model instead of normally distributed innovations, to account for potential serial correlation amongst the random errors. However, the SARIMA-GARCH model made no appreciable difference to the SARIMA forecast, and hence is not reported in detail.

## 5.3 Implications and Extensions

Of the plausible scenarios presented in Section 4.2.2, the final scenario is the most favoured. Importantly, the results of these scenarios do not indicate that optimising road networks and introducing advanced-tech vehicles will eliminate collisions entirely. Transport routes will contain a mixture of vehicle technology levels with distinct safety, avoidance, and communicative capabilities that will not entirely remove the level of risk on the road. In addition, the risk homeostasis theory suggests that drivers may increase their risk appetite in response to an absolute reduction in the level of risk exposure. As such, it remains to be seen whether the optimisation of vehicle technologies and road infrastructures will significantly and permanently reduce collision rates. Nevertheless, the resulting level of risk will lie at some level above zero, and therefore so too will the level of collision rates.

We highlight that the introduced and extended model can be further adjusted. In this study, we adjust the collision rate based on the number of registered vehicles on Irish roads. Further studies can develop the model to adjust instead for vehicle-miles travelled (VMT), which may more accurately reflect the risk exposure of vehicles on transport networks. We also make the assumptions that significant advancements in road safety follow a Gompertz distribution and are independent events that last for a fixed period of 36 months. However, these advancements may instead follow a different distribution. They may also be positively correlated – the positive effects from one advancement may lead to further advancements soon thereafter – and may be better suited to be modelled as part of a Markov-switching Process (Malyshkina *et al.* 2009). Furthermore, the period of accelerated safety may last any arbitrary length of time rather than a fixed 36-month period. The model may also be extended by including parameters to account for stochastic traffic volumes due to changing patterns in vehicle-ownership rates, changes in vehicle capabilities, and changes in transport infrastructure, which may affect collision rates in a non-linear manner.

The rate of collision reduction can also be adjusted to be a stochastic process, as in (Grzelak and Oosterlee 2011), rather than a constant. In addition, rare but significant spikes in collision rates, can be incorporated in to the model. In financial terms, these events are categorised as 'black swan' events. However, these events are also applicable to motor vehicle collision rates as emerging risks become prevalent and 'jumps' in collision rates become plausible. Examples of 'jump' events could include external societal events that disrupt travel patterns, such as the lockdowns caused by COVID-19. Another prominent possibility could stem from plausible cyber-risks, which are the highest perceived threat to highly-automated or autonomous vehicles (Claus *et al.* 2017). Cyber-risk describes the potential takeover of automated vehicles as a result of software or hardware vulnerabilities. Exploiting these vulnerabilities can lead to a malicious takeover of a fleet of vehicles. Although an example of this event is yet to occur, small-scale experiments have shown it to a viable threat (Murphy *et al.* 2019). This event has been described as a 'natural catastrophe' event for insurers (Pütz *et al.* 2019), and can lead to a large yet brief spike in collision rates. Regardless of future developments, the extended Heston model presented in this study represents a novel approach to collision frequency forecasting.

# 6  Conclusion

The Heston model is introduced in this study as a platform upon which to forecast the evolution of motor vehicle collision rates. We extend the conventional Heston model to make it fit for purpose for forecasting seasonal collision rates. Our extended model is primarily driven by three parameters – collision rate reduction targets, the variance in collision rates, and a binary 'switching' parameter that signals an upcoming period of accelerated safety. Further parameters guide the estimates and offer an avenue to include informed beliefs into the model regarding the changing dynamics of the process being modelled. The stochastic process approach we take in this study sidesteps the 'constant-parameter' and 'stability' assumptions that often affect contemporaneous models. The stochastic volatility model is beneficial to road safety research as it combines non-linear and stochastically-evolving parameters with informed beliefs about an uncertain future.

The application of the model showed closed affinity (over 95% accuracy) to observed collision rates between 2014-2018 and matched the annual seasonality of collision rates, demonstrating its forecasting ability over short-term intervals. The Extended Heston model also outperformed previously used forecasting methods such as the Vasicek and SARIMA models. Despite the utility that the Extended Heston model has in short-term forecasting, the predominant benefit the Extended Heston model provides is its capability for long-term forecasting. The flexibility within the parameters of the model means it can be adjusted to account for both previously-observed data and reasonable expectations about the evolution of road safety. As such, the Extended Heston model provides an avenue for analytically discussing the expected evolution of road safety through scenario analyses. To this end, we re-performed the analysis for time periods ahead – for 2019-2023 and for 2019-2044. The purpose of incorporating long-term forecasts was to explore a number of plausible scenarios for the evolution of collision rates. The scenario that assumes there will be a modest downward trend in annual rates, in conjunction with a reducing variance, point toward a significant reduction in rates over time. The model results in this scenario suggests an average fall in collision rates of 50% by 2044, with 75% of simulations suggesting a fall in collision rates from current levels. Although these results cannot be verified at present, the Extended Heston model can nevertheless serve as a valuable aid in determining the effectiveness of implemented policies or the forecasting of collision rates.

# 7  References

Anastasopoulos, P.C., Mannering, F.L., Shankar, V.N., Haddock, J.E., 2012a. A study of factors affecting highway accident rates using the random-parameters tobit model. Accident Analysis & Prevention 45, 628-633.

Anastasopoulos, P.C., Shankar, V.N., Haddock, J.E., Mannering, F.L., 2012b. A multivariate tobit analysis of highway accident-injury-severity rates. Accident Analysis & Prevention 45, 110-119.

Arbib, J., Seba, T., 2017. Rethinking transportation 2020-2030. RethinkX, May.

Arismendi, J.C., Back, J., Prokopczuk, M., Paschke, R., Rudolf, M., 2016. Seasonal stochastic volatility: Implications for the pricing of commodity options. Journal of Banking & Finance 66, 53-65.

Behnood, A., Mannering, F.L., 2016. An empirical assessment of the effects of economic recessions on pedestrian-injury crashes using mixed and latent-class models. Analytic methods in accident research 12, 1-17.

Bemmaor, A.C., Glady, N., 2012. Modeling purchasing behavior with sudden "death": A flexible customer lifetime model. Management Science 58 (5), 1012-1021.

Bergomi, L., 2015. Stochastic volatility modeling CRC press.

Bijleveld, F., Commandeur, J., Koopman, S.J., Montfort, K.V., 2010. Multivariate non-linear time series modelling of exposure and risk in road safety research. Journal of the Royal Statistical Society: Series C (Applied Statistics) 59 (1), 145-161.

Chen, C., Hu, J., Meng, Q., Zhang, Y., Year. Short-time traffic flow prediction with arima-garch model. In: Proceedings of the 2011 IEEE Intelligent Vehicles Symposium (IV), pp. 607-612.

Chen, S., Saeed, T.U., Labi, S., 2017. Impact of road-surface condition on rural highway safety: A multivariate random parameters negative binomial approach. Analytic methods in accident research 16, 75-89.

Cicchino, J.B., 2017. Effectiveness of forward collision warning and autonomous emergency braking systems in reducing front-to-rear crash rates. Accident Analysis & Prevention 99, 142-152.

Claus, S., Silk, N., Wiltshire, C., 2017. Potential impacts of autonomous vehicles on the uk insurance sector. Bank of England Quarterly Bulletin, Q1.

Clements, L.M., Kockelman, K.M., 2017. Economic effects of automated vehicles. Transportation Research Record 2606 (1), 106-114.

Cohen, T., Cavoli, C., 2019. Automated vehicles: Exploring possible consequences of government (non) intervention for congestion and accessibility. Transport reviews 39 (1), 129-151.

Commandeur, J.J., Bijleveld, F.D., Bergel-Hayat, R., Antoniou, C., Yannis, G., Papadimitriou, E., 2013. On statistical inference in time series analysis of the evolution of road safety. Accident Analysis & Prevention 60, 424-434.

Cox, J.C., Ingersoll Jr, J.E., Ross, S.A., 1985. A theory of the term structure of interest rates. Econometrica 53 (2), 385-408.

European Commission, 2018a. Connected & automated mobility: For a more competitive europe. Europe on the Move. Brussels, BE.

European Commission, 2018b. Safe mobility: A europe that protects. Europe on the Move Brussels, BE.

European Commission, 2019a. Eu road safety policy framework 2021-2030 - next steps towards "vision zero". European Commission, Brussels, BE.

European Commission, 2019b. Road safety: Commission welcomes agreement on new eu rules to help save lives. European Commission, Brussels, BE.

European Commission, 2020. 100 climate-neutral cities by 2030 - by and for the citizens. Luxembourg, LU.

European Transport Safety Council, 2020a. The impact of covid-19 lockdowns on road deaths in april 2020. Brussels, BE.

European Transport Safety Council, 2020b. Ranking eu progress on road safety. Brussels, BE.

Fagnant, D.J., Kockelman, K., 2015. Preparing a nation for autonomous vehicles: Opportunities, barriers and policy recommendations. Transportation Research Part A: Policy and Practice 77, 167-181.

Fountas, G., Anastasopoulos, P.C., Abdel-Aty, M., 2018. Analysis of accident injury-severities using a correlated random parameters ordered probit approach with time variant covariates. Analytic methods in accident research 18, 57-68.

Fountas, G., Fonzone, A., Gharavi, N., Rye, T., 2020. The joint effect of weather and lighting conditions on injury severities of single-vehicle accidents. Analytic Methods in Accident Research, 100124.

Fountas, G., Pantangi, S.S., Hulme, K.F., Anastasopoulos, P.C., 2019. The effects of driver fatigue, gender, and distracted driving on perceived and observed aggressive driving behavior: A correlated grouped random parameters bivariate probit approach. Analytic Methods in Accident Research 22, 100091.

Grzelak, L.A., Oosterlee, C.W., 2011. On the heston model with stochastic interest rates. SIAM Journal on Financial Mathematics 2 (1), 255-286.

Guo, J., Huang, W., Williams, B.M., 2014. Adaptive kalman filter approach for stochastic short-term traffic flow rate prediction and uncertainty quantification. Transportation Research Part C: Emerging Technologies 43, 50-64.

Heston, S.L., 1993. A closed-form solution for options with stochastic volatility with applications to bond and currency options. The review of financial studies 6 (2), 327-343.

Ho, C.Q., Hensher, D.A., Mulley, C., Wong, Y.Z., 2018. Potential uptake and willingness-to-pay for mobility as a service (maas): A stated choice study. Transportation Research Part A: Policy and Practice 117, 302-318.

Islam, M., Alnawmasi, N., Mannering, F., 2020. Unobserved heterogeneity and temporal instability in the analysis of work-zone crash-injury severities. Analytic Methods in Accident Research, 100130.

Jägerbrand, A.K., Sjöbergh, J., 2016. Effects of weather conditions, light conditions, and road lighting on vehicle speed. SpringerPlus 5 (1), 505.

Jannusch, T., Shannon, D., Völler, M., Murphy, F., Mullins, M., 2021. Cars and distraction: How to address the limits of dms and improve safety benefits using evidence from german young drivers (preprint). Technology in Society.

Johnson, C., 2017. Readiness of the road network for connected and autonomous vehicles. RAC Foundation: London, UK.

Kaltenhäuser, B., Werdich, K., Dandl, F., Bogenberger, K., 2020. Market development of autonomous driving in germany. Transportation Research Part A: Policy and Practice 132, 882-910.

Kröger, L., Kuhnimhof, T., Trommer, S., 2019. Does context matter? A comparative study modelling autonomous vehicle impact on travel behaviour for germany and the USA. Transportation research part A: policy and practice 122, 146-161.

Lassarre, S., 2001. Analysis of progress in road safety in ten european countries. Accident Analysis & Prevention 33 (6), 743-751.

Litman, T., 2020. Autonomous vehicle implementation predictions: Implications for transport planning.

Ma, L., Yan, X., Wei, C., Wang, J., 2016. Modeling the equivalent property damage only crash rate for road segments using the hurdle regression framework. Analytic methods in accident research 11, 48-61.

Ma, L., Yan, X., Weng, J., 2015. Modeling traffic crash rates of road segments through a lognormal hurdle framework with flexible scale parameter. Journal of Advanced Transportation 49 (8), 928-940.

Malyshkina, N.V., Mannering, F.L., Tarko, A.P., 2009. Markov switching negative binomial models: An application to vehicle accident frequencies. Accident Analysis & Prevention 41 (2), 217-226.

Mannering, F., 2018. Temporal instability and the analysis of highway accident data. Analytic methods in accident research 17, 1-13.

Mannering, F.L., Shankar, V., Bhat, C.R., 2016. Unobserved heterogeneity and the statistical analysis of highway accident data. Analytic methods in accident research 11, 1-16.

Meyer, J., Becker, H., Bösch, P.M., Axhausen, K.W., 2017. Autonomous vehicles: The next jump in accessibilities? Research in transportation economics 62, 80-91.

Murphy, F., Pütz, F., Mullins, M., Rohlfs, T., Wrana, D., Biermann, M., 2019. The impact of autonomous vehicle technologies on product recall risk. International Journal of Production Research 57 (20), 6264-6277.

Najaf, P., Thill, J.-C., Zhang, W., Fields, M.G., 2018. City-level urban form and traffic safety: A structural equation modeling analysis of direct and indirect effects. Journal of transport geography 69, 257-270.

Ohishi, K., Okamura, H., Dohi, T., 2009. Gompertz software reliability model: Estimation algorithm and empirical validation. Journal of Systems and software 82 (3), 535-543.

Oviedo-Trespalacios, O., Afghari, A.P., Haque, M.M., 2020. A hierarchical bayesian multivariate ordered model of distracted drivers' decision to initiate risk-compensating behaviour. Analytic methods in accident research 26, 100121.

Pütz, F., Murphy, F., Mullins, M., 2019. Driving to a future without accidents? Connected automated vehicles' impact on accident frequency and motor insurance risk. Environment Systems and Decisions 39 (4), 383-395.

Rajabzadeh, Y., Rezaie, A.H., Amindavar, H., 2017. Short-term traffic flow prediction using time-varying vasicek model. Transportation Research Part C: Emerging Technologies 74, 168-181.

Ramstedt, M., 2008. Alcohol and fatal accidents in the united states—a time series analysis for 1950–2002. Accident Analysis & Prevention 40 (4), 1273-1281.

Regev, S., Rolison, J.J., Moutari, S., 2018. Crash risk by driver age, gender, and time of day using a new exposure methodology. Journal of safety research 66, 131-140.

Road Safety Authority, 2019. Road casualties and collisions in ireland 2017 – tables provisional. Dublin, IE.

Shannon, D., Murphy, F., Mullins, M., Rizzi, L., 2020. Exploring the role of delta-v in influencing occupant injury severities–a mediation analysis approach to motor vehicle collisions. Accident Analysis & Prevention 142, 105577.

Transport Scotland, 2019. Reported road casualties scotland 2018: A national statistics publication for scotland. Edinburgh, UK.

Tsekeris, T., Stathopoulos, A., 2010. Short-term prediction of urban traffic variability: Stochastic volatility modeling approach. Journal of Transportation Engineering 136 (7), 606-613.

Us Dot, 2016. Smart city challenge: Lessons for building cities of the future. Washington, DC.

Us Dot, 2017. Beyond traffic 2045: Trends and choices. Washington, DC.

Vasicek, O., 1977. An equilibrium characterization of the term structure. Journal of financial economics 5 (2), 177-188.

Winston, C., Maheshri, V., Mannering, F., 2006. An exploration of the offset hypothesis using disaggregate data: The case of airbags and antilock brakes. Journal of Risk and Uncertainty 32 (2), 83-99.

World Health Organization, 2018. Global status report on road safety 2018 World Health Organization.

Xie, Y., Lord, D., Zhang, Y., 2007. Predicting motor vehicle collisions using bayesian neural network models: An empirical analysis. Accident Analysis & Prevention 39 (5), 922-933.

Yue, L., Abdel-Aty, M.A., Wu, Y., Farid, A., 2019. The practical effectiveness of advanced driver assistance systems at different roadway facilities: System limitation, adoption, and usage. IEEE Transactions on Intelligent Transportation Systems.

Zhang, X., Pang, Y., Cui, M., Stallones, L., Xiang, H., 2015. Forecasting mortality of road traffic injuries in china using seasonal autoregressive integrated moving average model. Annals of epidemiology 25 (2), 101-106.

Zhang, Y., Sun, R., Haghani, A., Zeng, X., 2013. Univariate volatility-based models for improving quality of travel time reliability forecasting. Transportation research record 2365 (1), 73-81.

# 8 Appendix
## 8.1 Appendix A: Models used for Forecasting

Eqn. 1 and Eqn. 2 in Section 3.1 demonstrate the conventional Heston Stochastic Volatility Model. The model calculates Markov Chain stepwise changes in the underlying asset price $S_t$ by assuming it satisfies the stochastic differential equation:

$$dS_t = \mu S_t dt + \sqrt{v_t} S_t dW_t^S \quad (A1)$$

In Eqn. A1, $v_t$, the instantaneous variance or squared volatility, is a Cox-Ingersoll-Ross (1985) process, and each change in $v_t$ is defined as

$$dv_t = \kappa(\theta - v_t)dt + \xi\sqrt{v_t}dW_t^v \quad (A2)$$

To make the Heston model fit for purpose for forecasting collision rates, a number of adjustments and amendments are included in Section 3.2, with the resulting Extended Heston model being:

$$dC_t = -\left(\mu G_t C_1 dt + \sqrt{v_t} C_1 dW_t^C\right) + \left(\overline{C_t^Y} \times A\sin(2\pi ft + \varphi)\right) \quad (A3)$$

$$dv_t = \kappa(\theta - v_t)dt + \xi\sqrt{v_t}dW_t^v \quad (A4)$$

$$G_{t\ldots t+s} = \begin{cases} \alpha, if\ G(T;b,\eta) > \text{unif}[0,1], \\ \quad\quad \alpha \in \text{unif}(\mathbb{N}[l,h]) \\ 1, if\ G(T;b,\eta) < \text{unif}[0,1] \end{cases} \quad (A5)$$

The adjustments can be summarised as:

- Adding an extra parameter ($G$) to account for the likelihood and extent of accelerated periods of road safety in future periods (Eqn. A5),
- Adding a sinusoidal adjustment to account for the seasonality in accident rates (Eqn. A3), and
- Changing the state-dependency in collision rate sizes from $C_t$ to $C_1$ (Eqn. A3), to instead rely on the assumption that stepwise changes in collision rates are a function of the initial collision rate rather than the prevailing collision rate.

In other to demonstrate the efficacy of the Extended Heston model, we compare the forecasting accuracy of the model against other models proposed in the literature. This includes ARIMA models (Ramstedt 2008), ARIMA-GARCH models (Chen *et al.* 2011, Guo *et al.* 2014), and the Vasicek model (Rajabzadeh *et al.* 2017). The results are provided in Section 5.1.

### 8.1.1 Vasicek Model for Short-term Interest Rates

The Vasicek (1977) model, like the Heston model, assumes that the rates being modelled satisfy a stochastic differential equation (Eqn. A6). In the Vasicek model, $\sigma$ is the instantaneous (constant) volatility, $\theta$ is the long-run interest rate, $r_t$ is the prevailing short-term rate, and $\kappa$ determines the speed at which $r_t$ reverts to $\theta$:

$$dr_t = \kappa(\theta - r_t)dt + \sigma dW_t \quad (A6)$$

Taken together, the $\kappa(\theta - r_t)dt$ represents a positive drift among stepwise changes, while the $\sigma dW_t$ represents normally-distributed innovations scaled to the size of the constant volatility $\sigma$. However, to ensure comparability against the Extended Heston model, the adjustments that are summarised in Appendix A are also afforded to the Vasicek model, where appropriate. This has the effect of transforming the Vasicek model in Eqn. A6 in to an adjusted Vasicek model (Eqn. A7, Eqn. A8):

$$dC_t = -(G_t \kappa(\theta - C_1)dt + \sigma dW_t) + (\overline{C_t^Y} \times A \sin(2\pi f t + \varphi)) \quad (A7)$$

$$G_{t \ldots t+s} = \begin{cases} \alpha, & if\ G(T; b, \eta) > \text{unif}[0,1], \\ \quad \alpha \in \text{unif}(\mathbb{N}[l, h]) \\ 1, & if\ G(T; b, \eta) < \text{unif}[0,1] \end{cases} \quad (A8)$$

The positive drift term $\kappa(\theta - C_1)dt$ is transformed to a negative drift to account for expected reductions over time. The negative drift is scaled by the 'acceleration' parameter $G$ (Eqn. A5). The state-dependency in collision rate sizes is again changed from $C_t$ to $C_1$. Finally, the sinusoidal adjustment is incorporated to account for the seasonality in collision rates. Figure A1 demonstrates the performance of the Adjusted Vasicek model in out-of-sample testing (forecasting 2014-2018 collision rates based on 2009-2013 parameters).

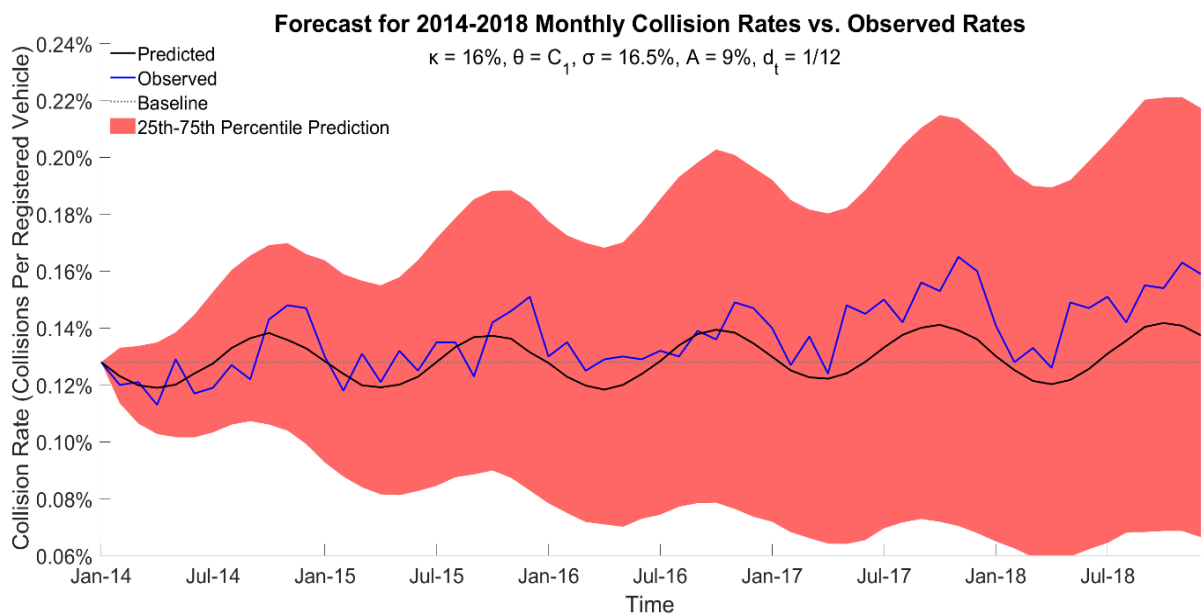

**Figure A1:** Observed vs. Predicted Values for Monthly Collision Rates from 2014-2018, starting from the January 2014 rate of 0.128%. Predictions were based on the Adjusted Vasicek model outlined in Appendix A, where the parameters were discerned from monthly collision rates covering 2009-2013 (Table B1, Table C1).

### 8.1.2 Seasonal Autoregressive Integrated Moving Average (SARIMA) model

Section 2 details how the collision rate process exhibits non-stationary elements; not only do the collision rates vary from month to month, but the process itself varies over different time periods. In order to capture the variability within the data, we additionally make use of an Autoregressive Integrated Moving Average (ARIMA) model. This approach has previously been utilised to forecast alcohol-related road fatalities in the United States (Ramstedt 2008).

The ARIMA model forecasts a time series process by assuming that future values can be determined using information gleaned from any combination of past values, past stepwise value changes, and past errors, respectively. The model is made up of autoregressive (AR), differencing (I), and moving average (MA) terms. Autoregressive (AR) terms are included to forecast the prevailing collision rate using a linear combination of trailing data points. Differencing (I) terms are included to account for the evidence found in Section 2 that the collision rate process itself is dynamic and subject to change over time. Therefore, rather than forecasting the specific position of the prevailing data point, we instead forecast the expected change for each prevailing data point relative to the previous observation. Finally, the moving average (MA) process believes information can be extracted from a linear combination of past values, similar to the AR process. However, rather than generating the linear combination using the past values themselves, the moving average process generates its linear combination using the prediction errors associated with past values.

Typically, given observations $y_t$, an ARIMA$(p,d,q)$ model with autoregressive lags $p$, difference term $d$ (I), and moving average lags $q$ can be stated as:

$$y'_t = c + \phi_1 y'_{t-1} + \cdots + \phi_p y'_{t-p} + \theta_1 \varepsilon_{t-1} + \cdots + \theta_q \varepsilon_{t-q} + \varepsilon_t \quad (\text{A9})$$

Where $y'_t$ denotes the 'differenced' observations, i.e. $y'_t = y_t - y_{t-1}$. In this generalised model, $\phi$ represents the weights attached to the autoregressive (AR) process and $\theta$ represents the weights attached to the moving average (MA) process. Furthermore, $p$ denotes the lookback period from which information relating to historical values is gleaned, while $q$ denotes the lookback period from which information relating to historical prediction errors is extracted. Finally, $c$ represents the intercept, while $\varepsilon_t$ represents normally-distributed innovations.

Given the extensive number of lags that can be used in ARIMA models, an alternative representation of Eqn. A9 is often provided through backshift notation, where $B^x y_t = y_{t-x}$. This transforms Eqn. A9 into:

$$y'_t = c + (\phi_1 B + \cdots + \phi_p B^p) y'_t + (1 + \theta_1 B + \cdots + \theta_q B^q) \varepsilon_t \quad (\text{A10})$$

Or

$$(1 - \phi_1 B - \cdots - \phi_p B^p) y'_t = c + (1 + \theta_1 B + \cdots + \theta_q B^q) \varepsilon_t \quad (\text{A11})$$

And given $y'_t = y_t - y_{t-1} = (1 - B) y_t$, we have:

$$(1 - \phi_1 B - \cdots - \phi_p B^p)(1 - B)^1 y_t = c + (1 + \theta_1 B + \cdots + \theta_q B^q) \varepsilon_t \quad (\text{A12})$$

However, Section 2 also outlines that there is a strong seasonal element to the data (Figure 4, Figure C1). Collision rates typically spike relative to the annual average during autumn, while there is a relative trough in collision rates during spring. Hence, we adopt the use of a SARIMA model. The SARIMA model is a variant of the ARIMA model that accounts for seasonality that is embedded in the data. The transforms the ARIMA$(p,d,q)$ model into an SARIMA$(p,d,q) \times (P,D,Q)_m$ model. Whereas the $p, d$ and $q$ values represent the lags associated with the non-seasonal portion of the model, the $P, D,$ and $Q$ lags describe the seasonal portion of the model. $m$ indicates the number of observations that describe each seasonal cycle; in our case, $m = 12$. SARIMA models have previously been utilised in road safety literature as a means of forecasting annual collision fatalities in China (Zhang *et al.* 2015).

Therefore, after accounting for seasonality cycles $m$, a SARIMA model with seasonal and non-seasonal differencing terms included ($d = D = 1$) takes the form:

$$(1 - \phi_1 B - \cdots - \phi_p B^p)(1 - \Phi_1 B - \cdots - \Phi_p B^P)(1 - B)^1 (1 - B^{12})^1 y_t = c + (1 + \theta_1 B + \cdots + \theta_q B^q)(1 + \Theta_1 B^{12} + \cdots + \Theta_q B^{12Q}) \varepsilon_t \quad (\text{A13})$$

The number of lags chosen for $p, q, P$ and $Q$ were based on a minimised Akaike Information Criterion (AIC) score. The lags were determined based on the SARIMA$(p,d,q) \times (P,D,Q)_m$ model that provided the closest affinity to 2009-2013 collision rates. The minimised AIC score was found with a SARIMA$(7,1,1) \times (1,1,2)_{12}$ model. This optimal fit model was then used to forecast 2014-2018 collision rates.

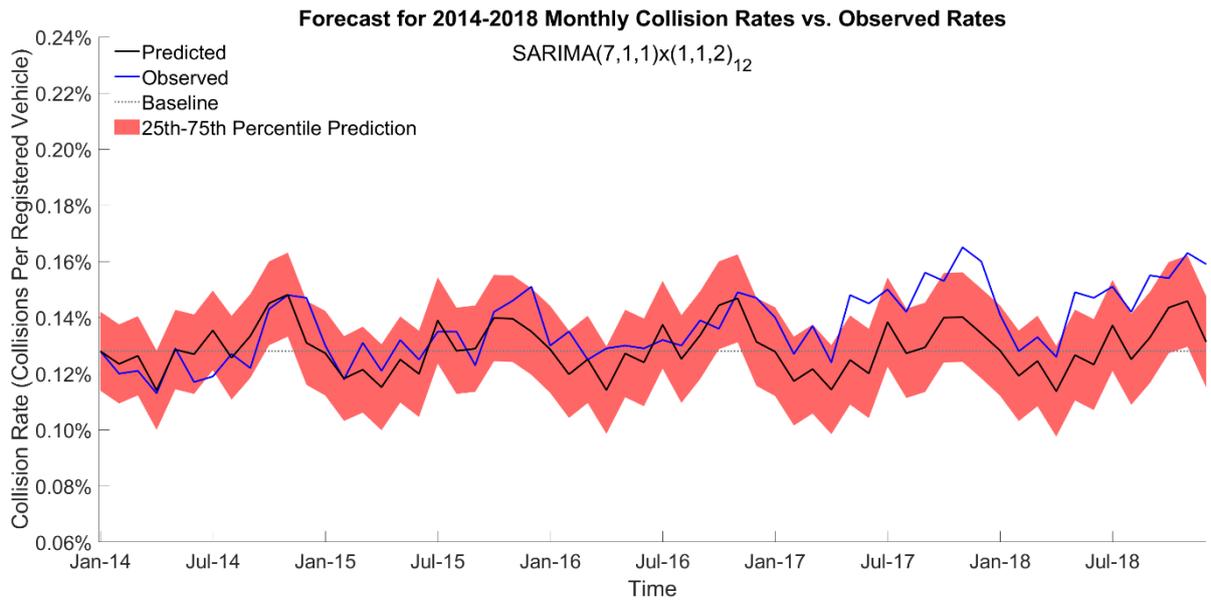

**Figure A2:** Observed vs. Predicted Values for Monthly Collision Rates from 2014-2018, starting from the January 2014 rate of 0.128%. Predictions were based on the SARIMA$(7,1,1) \times (1,1,2)_{12}$ model outlined in Appendix A, where the parameters were discerned from monthly collision rates covering 2009-2013 (Table B1, Table C1).

## 8.2 Appendix B: 2009-2013 Collision Rates and Summary Statistics

| | Month | Collisions | Collisions Per Registered Vehicle | Deviations from Average (Appendix C) | Collision Rate Log-differences | Yearly Volatility | Volatility Log-differences |
|---|---|---|---|---|---|---|---|
| Min | | 1,799 | 0.075% | -18.83% | -30.36% | | |
| Max | | 2,805 | 0.116% | 24.24% | 28.04% | | |
| Mean | | 2,226 | 0.091% | 0 | 0.18% | | |
| Std. Dev | | 220.47 | 0.009% | 9.78% | 11.71% | | |
| 2009 (Registered Vehicles: 2,468,000) | Jan | 2,278 | 0.092% | 3.17% | - | 45.33% | - |
| | Feb | 1,919 | 0.078% | -13.09% | -17.15% | | |
| | Mar | 1,965 | 0.080% | -11.00% | 2.37% | | |
| | Apr | 1,895 | 0.077% | -14.17% | -3.63% | | |
| | May | 2,157 | 0.087% | -2.31% | 12.95% | | |
| | June | 1,934 | 0.078% | -12.41% | -10.91% | | |
| | July | 2,560 | 0.104% | 15.95% | 28.04% | | |
| | Aug | 2,275 | 0.092% | 3.04% | -11.80% | | |
| | Sep | 2,145 | 0.087% | -2.85% | -5.88% | | |
| | Oct | 2,431 | 0.099% | 10.10% | 12.52% | | |
| | Nov | 2,455 | 0.099% | 11.19% | 0.98% | | |
| | Dec | 2,481 | 0.101% | 12.37% | 1.05% | | |
| 2010 (Registered Vehicles: 2,416,000) | Jan | 2,217 | 0.092% | -1.78% | -9.12% | 55.82% | 20.82% |
| | Feb | 2,044 | 0.085% | -9.44% | -8.12% | | |
| | Mar | 2,208 | 0.091% | -2.17% | 7.72% | | |
| | Apr | 1,832 | 0.076% | -18.83% | -18.67% | | |
| | May | 2,282 | 0.094% | 1.10% | 21.96% | | |
| | June | 2,135 | 0.088% | -5.41% | -6.66% | | |
| | July | 2,337 | 0.097% | 3.54% | 9.04% | | |
| | Aug | 1,997 | 0.083% | -11.52% | -15.72% | | |
| | Sep | 2,442 | 0.101% | 8.19% | 20.12% | | |
| | Oct | 2,745 | 0.114% | 21.62% | 11.70% | | |
| | Nov | 2,788 | 0.115% | 23.52% | 1.55% | | |
| | Dec | 2,058 | 0.085% | -8.82% | -30.36% | | |
| 2011 (Registered Vehicles: 2,425,000) | Jan | 2,244 | 0.093% | -0.61% | 8.28% | 41.10% | -30.70% |
| | Feb | 2,078 | 0.086% | -7.96% | -7.69% | | |
| | Mar | 2,071 | 0.085% | -8.27% | -0.34% | | |
| | Apr | 1,907 | 0.079% | -15.54% | -8.25% | | |
| | May | 2,054 | 0.085% | -9.02% | 7.43% | | |
| | June | 2,288 | 0.094% | 1.34% | 10.79% | | |
| | July | 2,805 | 0.116% | 24.24% | 20.37% | | |
| | Aug | 2,255 | 0.093% | -0.12% | -21.83% | | |
| | Sep | 2,245 | 0.093% | -0.56% | -0.44% | | |
| | Oct | 2,437 | 0.100% | 7.94% | 8.21% | | |
| | Nov | 2,209 | 0.091% | -2.16% | -9.82% | | |
| | Dec | 2,500 | 0.103% | 10.73% | 12.38% | | |
| 2012 (Registered Vehicles: 2,403,000) | Jan | 2,092 | 0.087% | -4.08% | -16.91% | 32.56% | -23.19% |
| | Feb | 1,799 | 0.075% | -17.51% | -15.09% | | |
| | Mar | 2,156 | 0.090% | -1.14% | 18.10% | | |
| | Apr | 2,151 | 0.090% | -1.37% | -0.23% | | |
| | May | 2,038 | 0.085% | -6.55% | -5.40% | | |
| | June | 2,191 | 0.091% | 0.46% | 7.24% | | |
| | July | 2,213 | 0.092% | 1.47% | 1.00% | | |
| | Aug | 2,286 | 0.095% | 4.82% | 3.25% | | |
| | Sep | 2,261 | 0.094% | 3.67% | -1.10% | | |
| | Oct | 2,361 | 0.098% | 8.26% | 4.33% | | |
| | Nov | 2,279 | 0.095% | 4.50% | -3.53% | | |
| | Dec | 2,344 | 0.098% | 7.48% | 2.81% | | |
| 2013 (Registered Vehicles: 2,483,000) | Jan | 2,277 | 0.092% | 2.30% | -6.17% | 29.67% | -9.32% |
| | Feb | 1,932 | 0.078% | -13.20% | -16.43% | | |
| | Mar | 2,099 | 0.085% | -5.70% | 8.29% | | |
| | Apr | 1,979 | 0.080% | -11.09% | -5.89% | | |
| | May | 2,187 | 0.088% | -1.74% | 9.99% | | |
| | June | 2,088 | 0.084% | -6.19% | -4.63% | | |
| | July | 2,304 | 0.093% | 3.51% | 9.84% | | |
| | Aug | 2,215 | 0.089% | -0.49% | -3.94% | | |
| | Sep | 2,259 | 0.091% | 1.49% | 1.97% | | |
| | Oct | 2,464 | 0.099% | 10.70% | 8.69% | | |
| | Nov | 2,355 | 0.095% | 5.80% | -4.52% | | |
| | Dec | 2,551 | 0.103% | 14.61% | 7.99% | | |

| Model Parameters | January '14 Rate[8] | Amplitude (Appendix C) | 5-year Volatility | 5-year Volatility of Volatility |
|---|---|---|---|---|
| | 0.128% | 9% | 40.57% | 22.74% |

**Table B1:** Summary statistics and discerning model parameters from a time series of monthly collision data, 2009-2013.

## 8.3 Appendix C: Seasonality Associated with 2009-2013 Collision Rates

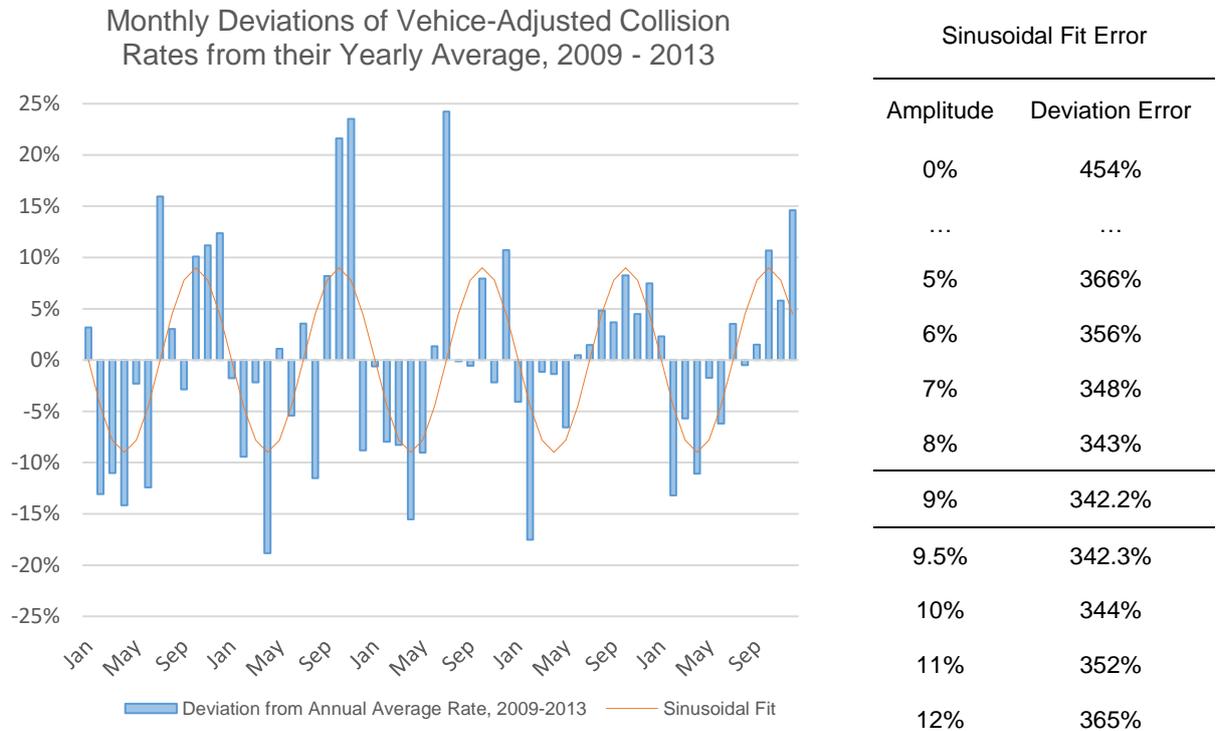

Sinusoidal Fit Error

| Amplitude | Deviation Error |
|---|---|
| 0% | 454% |
| … | … |
| 5% | 366% |
| 6% | 356% |
| 7% | 348% |
| 8% | 343% |
| 9% | 342.2% |
| 9.5% | 342.3% |
| 10% | 344% |
| 11% | 352% |
| 12% | 365% |

**Figure C1:** Monthly collision rates are plotted relative to the average rate of each year. The deviations of each month from their annual average follows a set pattern (lower than average the first six months, higher than average the latter six months) that can be reasoned as following that of a sine wave that begins halfway through its cycle.

**Table C1:** Absolute differences between sine wave cycles and monthly deviations in collision rates from their yearly average for different amplitudes.

---

[8] The move from manual reports to electronic recordings of vehicles in January 2014 dramatically increased the collision rate (Figure 2). It is believed that the January 2014 collision rate represented a 'new norm' that was not reflected in the December 2013 rate. Therefore, we started from the January 2014 level, rather than December 2013.